\documentclass[letterpaper,twocolumn,10pt]{article}
\usepackage{usenix-2020-09}

\usepackage{amsmath} 
\usepackage[pdftex]{graphicx} 
\DeclareGraphicsExtensions{.pdf,.eps,.png,.jpg}
\graphicspath{{images/}{Paper/images/}} 
\usepackage{booktabs} 
\usepackage[inline]{enumitem} 
\usepackage{balance} 
\usepackage{wasysym}
\usepackage{subcaption}
\usepackage{censor}
\usepackage{float}
\usepackage[compact]{titlesec}

\usepackage{colortbl}
\usepackage{tcolorbox}

\usepackage{tikz}
\usetikzlibrary{shapes.geometric, arrows, positioning, calc, backgrounds}

\usepackage[font=small,format=plain,labelfont=bf,textfont=it]{caption}

\renewcommand{\paragraph}[1]{\vspace*{0.05in}\noindent{\bf #1}}

\makeatletter
\def\input@path{{./sections/}{./misc/}{./Paper/sections/}{./Paper/misc/}}
\makeatother

\begin{document}

\title{User Awareness and Behaviors Concerning \\ Encrypted DNS Settings in Web Browsers}


\author{
    Alexandra Nisenoff~\footnotemark[2]~\footnotemark[1]~, Ranya
    Sharma\footnotemark[1], Nick Feamster\footnotemark[1] \\
    {\normalsize \footnotemark[1]~University of
    Chicago~~~\footnotemark[2]~Carnegie Mellon University} 
}

\maketitle

Recent developments to encrypt the Domain Name System (DNS) have resulted in
major browser and operating system vendors deploying encrypted DNS
functionality, often enabling various configurations and settings by default.
In many cases, default encrypted DNS settings have implications for
performance and privacy; for example, Firefox's default DNS setting sends all
of a user's DNS queries to Cloudflare, potentially introducing new privacy
vulnerabilities. In this paper, we confirm that most users are unaware of these
developments---with respect to the rollout of these new technologies, the
changes in default settings, and the ability to customize encrypted DNS
configuration to balance user preferences between privacy and performance.
Our findings suggest several important implications for the designers of
interfaces for encrypted DNS functionality in both browsers and operating
systems, to help improve user awareness concerning these settings, and to
ensure that users retain the ability to make choices that allow them to
balance tradeoffs concerning DNS privacy and performance.

\section{Introduction}\label{sec:intro}


The Domain Name System (DNS) is an Internet protocol that maps human-readable
domain names to Internet Protocol (IP) addresses.  Conventionally, DNS queries
have been unencrypted, leaving both queries and responses vulnerable to
passive eavesdropping and active manipulation.  In recent years, major browser
and operating system vendors have begun to deploy protocols such as
DNS-over-TLS (DoT) and DNS-over-HTTPS (DoH), which encrypt queries and DNS.
With the emergence of this functionality, browsers now have different default
settings for encrypted DNS, as well as various configuration options to allow
users to customize how their browser performs DNS lookups.  As is the case
with browser vendors, the ``Private DNS'' mode on Android uses DNS-over-TLS
(DoT) to encrypt DNS queries and responses in between the client and the DNS
resolver.  Encrypted DNS settings are implemented in subtly different ways,
and the settings have a variety of defaults.

These differences lead to the setting behaving in ways with implications users may not understand. 
Getting users to make informed choices of encrypted DNS settings is difficult
due to their technical nature~\cite{Knodel-22-DNSPrivacyVs}.
The obscurity of these settings may be further
exacerbated by the lack of information provided to users when they modify
these settings. 
Nevertheless, the implications of these settings are
significant: depending on how users configure encrypted DNS, for example, all
of a user's DNS queries may be sent to a single DNS provider, such as Google
or Cloudflare. Occasionally, such configurations are even changed without a
user's awareness, with browser vendors pushing changes to default settings
with browser version and operating system upgrades. Although users can
typically change these settings, doing so requires an awareness that
configuring encrypted DNS settings is possible, knowledge about different
configuration choices, and the ability to change these settings. 
It is important to give careful consideration to default settings and interfaces in light of the fact that they often go unmodified by users~\cite{weinmann2016digital}.  
Furthermore, previous research has observed that encrypted DNS is what Clark {\em et al.} have described as an Internet ``tussle space''~\cite{clark2002tussle}, where various Internet stakeholders may vie for control through technical protocol design.
Indeed, Hounsel et al. have identified encrypted DNS as a tussle
space~\cite{hounsel2021designing}, and highlighted the importance of designing the encrypted DNS infrastructure and user interfaces in ways that {\em preserve
user choice}. 
Clark explains the profound importance of maintaining user choice in Internet
protocols as follows, ``It is important that protocols be designed in such a
way that all the parties to an interaction have the ability to express preference about which
other parties they interact with...it matters if the consequence of choice is
visible.''~\cite{clark2002tussle}.

The design choices in the encrypted DNS ecosystem have implications for market consolidation and power, corporate visibility into and control over user data, as well as user privacy on the Internet.
Given the significant implications of these settings, and users' ability to change them, it is critical to understand how users understand these settings and interact with them, which is inherently dependent on how they are presented by vendors.

In this paper, we study encrypted DNS settings in Brave, Chrome, Edge,
Firefox, Opera, and the Android mobile operating system. (We exclude Safari
because, both at the time of our study and as of February 2023, it does not offer an encrypted DNS setting.)
Previous small-scale (approximately ten users) research~\cite{nisenoff-21-PrivateDNS} included a preliminary study of these questions involving Android's Private DNS setting. In this study, we expand on these results to give
a larger scale look at how users interact with these settings in the context
of browsers.
We study the following questions:
\begin{itemize}
\itemsep=-2pt
    \item Do users know about and trust encrypted DNS and its stakeholders?
        (Section~\ref{sec:awareness})
    \item What encrypted DNS settings do users have enabled in their browsers
        and phones? (Section~\ref{sec:settings})
    \item When shown encrypted DNS settings for different browsers, which settings do
        users choose, and why? (Section~\ref{sec:selection})
    \item When the technical aspects of these systems are explained to users,
        do their preferences for settings change?
        (Section~\ref{sec:technical})
\end{itemize}

\paragraph{Summary of findings.}
Many users have heard of DNS but do not know what it does. Even users who
believed they knew about DNS's function were often incorrect in describing it.
Users typically had the default settings enabled in their own browsers. When
users deviated from the default, they tended to select either Cloudflare or
Google as their recursive resolver or simply disabled the setting. In many
cases, the default setting opportunistically encrypts their DNS queries. In
mobile devices, a surprising number of users had the ``Private DNS'' setting
disabled, despite that not being the default setting. 

More than 70\% of participants continued to use the default encrypted DNS
settings, and this tendency varied across different interfaces. Moreover, the
way the settings were described played a role in users' decision making. 37\%
of participants chose to change their encrypted DNS settings after receiving
an explanation of DNS and encrypted DNS. Although the defaults remained
popular, users did select a variety of settings after receiving more information. 
The setting to enter a custom resolver was not often selected, but when
participants did attempt to specify a custom resolver, none of the participants
entered text that would have functioned properly. Participants seemed to
understand the settings better after an explanation, yet they still had
problems understanding the differences in functionality and privacy guarantees
among the different resolvers.

\paragraph{Summary of recommendations.}
These results lead to a number of practical recommendations for designers of encrypted DNS
interfaces, standards bodies, and other policymakers, which we detail in
Section~\ref{sec:disc}. In particular, users need a basic primer on DNS,
its associated privacy risks, the guarantees that encrypted DNS can (and
cannot) provide, and an intuitive way to understand the implications of the
choices for different recursive resolvers. Users also need easier ways
to customize the choices for trusted recursive resolvers, once the
implications of these choices are clear. Future work can and should expand on
our initial findings to explore interfaces for configuring encrypted DNS that
provide users choices for configuration, as well as a clear
understanding of the implications of these choices.

\section{Background and Related Work}\label{sec:back}

This section surveys background and related work on encrypted DNS,
including the basic operation and privacy risks of DNS queries, the
functionality of encrypted DNS, and the history of the introduction of
encrypted DNS into modern browsers and operating systems.  We also survey past
related work concerning users' awareness concerning privacy settings in
browsers, including encrypted DNS.

\subsection{Background: DNS and Encrypted DNS}

Most Internet connections are preceded by a Domain Name System (DNS) lookup,
which maps a human-readable name to an Internet protocol (IP) address that a
client can use to ultimately connect to a remote destination or service.  DNS
queries and responses have historically been unencrypted, and they may contain
sensitive information including information about the site that a user is
visiting. Previous research has shown that observing a user's DNS queries can
allow users to be tracked across multiple
websites~\cite{bortzmeyer2015dns,Moura2020Centralized,herrmann2010analyzing}.
Beyond tracking users online, DNS traffic can also be used to infer what
``smart'' Internet of things (IoT) devices are present in a home, and may even
expose information on how people use
them~\cite{apthorpe2017EncIoTTraffic,apthorpe2017smart,le2019policy}.  Some
Internet service providers have logged DNS queries to their resolvers and
shared them with third parties~\cite{DNSprivProb}.  Because these queries are
typically made in plaintext, anyone who can observe a user's network traffic
could see the contents of these queries.  Outside of the privacy implications
of the DNS, there are also potential concerns with integrity, where the
responses to DNS queries may be modified resulting in users receiving the
incorrect IP for a website; a censor may be able to implement this type of
manipulation to block access to a legitimate website or redirect users to
another website entirely~\cite{pearce2017global,bortzmeyer2015dns}. 

To mitigate some of the security and privacy issues with DNS, Internet
operators and vendors have introduced DoT and DoH, both of which send DNS
queries using encrypted protocols, with subtle differences in their
implementations~\cite{hu2016DoTSpec, hoffman2018DoHSpec}.  Encrypting DNS
queries and responses prevents passive eavesdroppers from observing the
content of users' DNS queries.  DoT sends queries over a Transport Layer
Security (TLS) connection using port 853, while DoH uses HTTPS rather than TLS
for the transport protocol, typically on port 443.  Because DoT uses a
dedicated port, it is easier to monitor and detect (and potentially block); on
the other hand, DoH uses port 443 for transport and thus DoH traffic tends to
be more difficult to identify, particularly when combined with other HTTPS
traffic~\cite{HvT}.  Although both of these protocols encrypt DNS queries and
responses, they are still susceptible to various
attacks, ranging from downgrade attacks to traffic analysis-based inference
attacks~\cite{unpaddedDoH,Siby2020EncDNSTraffic,Huang2020DoHDowngrade,Houser2019DoTLeakage,vekshin2020dohML}.

Using DoH and DoT can provide users with many benefits, but there are
downsides to using these protocols as well.  Many existing systems such as
parental filtering, safe search, or malware detection tools rely on access to
the content of DNS queries. However, encrypting those queries while using DoH
or DoT can sometimes inhibit the functioning of these
systems~\cite{DNSprivProb,ff1,lemos2013got, Hounsel2021Tussle,
Lyu2022SurveyEncDNS}.  Similarly, because Internet service providers (ISPs) can be required by government
to block access to certain illegal content, implementation of encrypted DNS
may in some cases prevent ISPs from complying with these laws~\cite{hvt2}.  On
the privacy front, because DoH and DoT need to communicate with resolvers that
support these protocols, encrypted DNS may result in users' queries being sent
to fewer resolvers, allowing them to see more of a user's online behavior, given that fewer entities can observe more of a user's DNS queries~\cite{borgolte2019ecosystem,
Hounsel2021Tussle}. This consolidation is evident in the growing consolidation of
hosting authoritative DNS resolution in
general~\cite{Wang2021Consolidation}.

Although DoH and DoT provide security for the queries while they are in transit,
these protocols do not prevent the operators of these DNS resolvers from learning about users'
queries or ensuring that the DNS resolver returns the correct response. Other
proposed improvements to DNS designed to address these issues include:
Oblivious DNS~\cite{ODNS}, Oblivious DNS over HTTPS~\cite{ODoH},
secure DNS~(DNSSEC)~\cite{eastlake1999DNSSEC,arends2005protocol}, query name~(QNAME)
minimization~\cite{bortzmeyer2016QNAMEMin}, and dividing queries across
multiple resolvers~\cite{kresolver2020, Hounsel2021DistributingQueries}.

\subsection{User Awareness of Encryption\\ and Privacy Settings}

This study focuses on user awareness concerning encrypted DNS and its
configuration through common interfaces, but many other studies have
investigated how to communicate security and privacy concepts to users.  While
usually less buried in setting menus, private browsing modes are included in
browsers and do not store browsing history, cookies, or temporary files across
sessions.  Research has shown that users have many misconceptions about what
these setting actually do.  Much like encrypted DNS settings, each browser
provides a unique description of each setting, which both play a role in what
protections the users think the setting provides and have been shown to be
insufficient in correcting common misconceptions about what the settings
do~\cite{wu18privatebrowsing,gao14privatebrowsing}.  Other research into the
communication of security risks to users have covered topics such as SSL
warnings~\cite{devdatta13warningland,felt15improveSSLwarnings,
sunshine09crywolf}, visual
icons~\cite{Felt16ConnectionIndicators,Habib21CCPAIcon}, privacy
policies~\cite{Wilson16AnnotatePrivacyPolicies,tsai11purchasingbehavior},
privacy notices~\cite{Schaub15EffectivePrivacyNotices}, social media
privacy settings~\cite{Liu2011FbPrivSettings}, and cookie consent interfaces~\cite{habib2022_okaywhatever}.  In the realm of encrypted DNS
settings, one small-scale exploratory study found that users do not understand
the impact of different setting options in the PrivateDNS setting on Android
and that most users would initially choose the default options, but when given more information on the setting, some users did choose to modify their
choice~\cite{nisenoff-21-PrivateDNS}.  In this paper, we expand on those ideas
and explore encrypted DNS settings in the browsers on a larger scale.

Several studies have shown that users make incorrect assumptions about the
security guarantees of encryption, have doubts about the protection from
adversaries, misunderstand phrases like end-to-end encryption, or have
incorrect mental models relative to protection provided in different
contexts~\cite{Wu18EncMentalModels,
Abu-Salma18e2eMentalModels,Stransky2021visEnc, Gerber18JohnnyCanEncrypt, Dechand19EncDontTrust,
Abu-Salma17ObstaclestoAdoption}.  Technical jargon, paired with inconsistent
terminology, can also make tools that use encryption even more difficult for the
average user to use correctly~\cite{abu2017securityblanket}.  Beyond how users
react to and interact with settings, it is also helpful to understand users'
perception of encryption in other contexts.  If individuals use
encryption tools incorrectly, they can have a false sense of security or get themselves
into situations where they can no longer perform the tasks they were
originally attempting to do.

\section{Interfaces for Configuring Encrypted DNS}
\label{sec:interface-analysis}
Vendors have increasingly added support for encrypted DNS, including Web browsers and mobile devices. 
In this section, we survey the current state of the interfaces for configuring encrypted DNS in common Web browsers.
We focus in particular on the interfaces for configuring encrypted DNS in five
different popular browsers---Chrome, Brave, Firefox, Microsoft Edge, and
Opera---and one mobile operating system, Android. We focus on how these
settings are presented to users in the United States; these interfaces may
differ in other regions or countries.

\begin{table*}[t]
    \centering
    \begin{scriptsize}
    \resizebox{\linewidth}{!}{
    \begin{tabular}{ l|ccc|c}
    & \multicolumn{3}{c|}{\textbf{Browsers}} & \textbf{Mobile} \\ \hline

    \textbf{\begin{tabular}[c]{c}Platform\end{tabular}} & \textbf{Chromium} & \textbf{Firefox} & \textbf{Opera} & \textbf{Android} \\
     \hline
    \textbf{\begin{tabular}[c]{c}Version Where Introduced\footnotemark[1]\end{tabular}} & {Brave 1.7, Chrome 83, Edge 86} & {Firefox 73} & {Opera 65 Beta} & {Android 9 Pie}  \\
       
       \hline
        \textbf{\begin{tabular}[c]{c}Setting Name\end{tabular}}  & \shortstack{Secure DNS} & \shortstack{DNS over HTTPS} & \shortstack{DNS-over-HTTPS} & \shortstack{Private DNS} \\
        \hline
        \textbf{\begin{tabular}[c]{c}Protocol\end{tabular}} & DoH & DoH & DoH & DoT\\
        \hline
        \textbf{\begin{tabular}[c]{c}Default\end{tabular}} & Opportunistic  & Cloudflare & Disabled & Opportunistic \\

        \hline
        \textbf{\shortstack{Support for Opportunistic\\use of Encrypted DNS}} & \newmoon  & \fullmoon & \fullmoon & \newmoon\\

        \hline
        \textbf{\shortstack{Warning for Malformed\\Custom DNS Resolver URL}} & \newmoon  & \newmoon & \fullmoon & \fullmoon\\

        \hline
        \textbf{\shortstack{Links to Privacy Polices\\for Resolvers Shown to Users}} & \newmoon  & \fullmoon & \fullmoon & No resolvers shown \\
      
        \bottomrule
        
    \end{tabular}}
    \end{scriptsize}
    \caption{Summary of encrypted DNS interfaces.}\label{tab:interfaces}
\end{table*}
\footnotetext[1]{This was a best effort attempt to identify the first version where the setting appeared, based on news articles, GitHub issues, and release notes.}
\begin{table}[t]
    \begin{scriptsize}
    \centering
    \begin{tabular}{l|ccccc}
    \hline
    Browser    & Cloudflare & CleanBrowsing & Google    & NextDNS   & OpenDNS   \\
    \hline
    Chrome     & \newmoon  & \newmoon     & \newmoon & \newmoon & \newmoon  \\ 
    Firefox    & \newmoon  & \fullmoon    & \fullmoon& \newmoon & \fullmoon\\
        Edge\footnotemark[2]       & \newmoon  & \newmoon     & \newmoon & \newmoon & \newmoon   \\ 
        Opera\footnotemark[3]      & \newmoon  & \fullmoon    & \newmoon & \fullmoon& \fullmoon \\
    Brave\footnotemark[2]      & \newmoon  & \newmoon     & \newmoon & \newmoon & \newmoon    \\
    \bottomrule
    \end{tabular}
    \end{scriptsize}
    \caption{Resolvers listed in encrypted DNS settings interfaces by different browser vendors.}
    \label{tab:SupportedResolvers}
\end{table}

Figure~\ref{fig:enc-dns} in the Appendix shows the encrypted DNS setting
interfaces for
Brave, Chrome, Edge, Firefox, and Opera.  Table~\ref{tab:interfaces} and
Table~\ref{tab:SupportedResolvers} have detailed descriptions of the encrypted
DNS interfaces and the resolvers that they support.  Encrypted DNS interfaces
fall into two categories: (1)~Chromium-based interfaces (Brave, Chrome, and
Edge) and (2)~Firefox/Opera.  Chromium-based browsers refer to encrypted DNS
as ``secure DNS'' and default to opportunistically using the user's default
DNS resolver, while falling back to sending unencrypted DNS queries if the
encrypted DNS resolver is unavailable.  Edge provides many of the same options
for settings to users, but does so in a slightly different way: When a user
selects a resolver from the drop down menu, Edge automatically fills the text
box for selecting a custom resolver with the resolver's URL allowing it to be
edited by the user.

In contrast to Chromium-based browsers, Firefox and Opera refer to the
encrypted DNS setting as ``DNS-over-HTTPS'' and do not provide support for an
opportunistic mode, where the browser will encrypt queries if the DNS resolver they were already using supports DoH, as Chromium browsers do.  Firefox's default behavior is to
use encrypted DNS with Cloudflare as the default resolver; in contrast, Opera
disables encrypted DNS by default. When encrypted DNS is enabled, Opera uses
Cloudflare as the default resolver.  Only Opera explicitly mentions that DoH
uses third-party services; rather than falling back to unencrypted queries, it
alerts users that a page was inaccessible, mentioning that the DNS-over-HTTPS
setting may be to blame.  As far as selection options of resolvers shown to users, Mozilla operates a Trusted Recursive Resolver (TRR) program, which ensures that DoH
providers recommended by Firefox best protect privacy by not over-collecting
and sharing data~\cite{ff-trr}. 
Some of the resolvers shown in these interfaces offer additional features such as blocking malware or adult content.
\footnotetext[2]{After the survey was distributed Quad9 was removed from the list of resolvers in Brave \& Edge.}
\footnotetext[3]{Opera offers multiple versions of the Cloudflare resolver, including versions that block adult content and malware.}
\addtocounter{footnote}{3}

``Private DNS'' in Android is the only mobile operating system that we include
in our analysis.  Unlike the browser-based settings, ``Private DNS'' supports
DoT rather than DoH.  By default, DoT encrypts DNS queries to whichever
resolver the user has selected; Android's private DNS will fall back to
unencrypted queries if DoT fails.  In contrast to the settings in browsers,
Private DNS does not give the user any suggestions of resolvers, forcing the
user to input their own URL for a resolver.  This mode will not fall back to
unencrypted queries and can cause web pages or resources to become unavailable
if, for example, the user inputs an invalid URL.

\section{Method}\label{sec:method}


To learn more about participants' understanding of encrypted DNS settings in
browsers, we designed a two-part survey.  In the first part of the survey, we
asked participants about their usage of different browsers. Based on the
answers from the first part of the survey, the second part of the survey then
asked users to interact with a high-fidelity interface for encrypted DNS
settings designed to resemble their browser. We also asked users about encrypted DNS
settings in their own browser.  In this section, we first describe the survey design and
recruitment methods; we then discuss the limitations of our survey design, as
well as the ethical considerations associated with our survey design.  This
study was approved by our university's Institutional Review Board (IRB).

\subsection{Study Design}\label{sec:method:survey_design}

\begin{figure}[t] 
    \centering \label{fig:method-diagram}
    \resizebox{\columnwidth}{!}{ \tikzstyle{decision} = [diamond, draw, fill=blue!20, 
    text width=4.5em, text badly centered, node distance=3cm, inner sep=0pt]

\tikzstyle{block} = [rectangle, draw, fill=blue!20, 
    text width=15em, text centered, rounded corners, minimum height=2em]

\tikzstyle{screenblock} = [rectangle, draw, fill=blue!20, 
    text width=6em, text centered, rounded corners, minimum height=2em]

\tikzstyle{smallblock} = [rectangle, draw, fill=blue!20, 
    text width=5em, text centered, rounded corners, minimum height=2em]

\tikzstyle{line} = [draw, -latex', pos=0.5]

\begin{tikzpicture}[align=center, auto]
\linespread{1}
    \node [block,text width=8em] (usage) {\textbf{Collect information on browser usage from screening survey}};

\node [screenblock, right=.75cm of usage] (opera) {\textbf{Participant uses Opera at least once a day}};

\node [smallblock, below=.75cm of opera, fill=red!20] (operaassign) {\textbf{Assigned to Opera}\\n=24};

\node [screenblock, right=.75cm of opera] (brave) {\textbf{Participant uses Brave at least once a day}};
\node [smallblock, below=.75cm of brave, fill=orange!20] (braveassign) {\textbf{Assigned to Brave}\\n=60};

\node [screenblock, right=.75cm of brave] (edge) {\textbf{Participant uses Edge at least once a day}};

\node [smallblock, below=.75cm of edge, fill=green!20] (edgeassign) {\textbf{Assigned to Edge}\\n=110};

\node [screenblock, right=.75cm of edge] (ff) {\textbf{Participant uses Firefox at least once a day}};
\node [smallblock, below=.75cm of ff, fill=yellow!20] (ffassign) {\textbf{Assigned to Firefox}\\n=126};

\node [screenblock, right=.75cm of ff] (chrome) {\textbf{Participant uses Chrome at least once a day}};
\node [smallblock, below=.75cm of chrome, fill=violet!20] (chromeassign) {\textbf{Assigned to Chrome}\\n=425};

\node [screenblock, right=.75cm of chrome, fill=lightgray!20] (discard) {\textbf{Do not assign user to any group}};

\path [line] (usage) -- (opera);

\path [line] (opera) -- (operaassign);
\path [line] (opera) -- node {yes}(operaassign);
\path [line] (opera) -- node {no}(brave);

\path [line] (brave) -- node {yes}(braveassign);
\path [line] (brave) -- node {no}(edge);

\path [line] (edge) -- node {yes}(edgeassign);
\path [line] (edge) -- node {no}(ff);

\path [line] (ff) -- node {yes}(ffassign);
\path [line] (ff) -- node {no}(chrome);

\path [line] (chrome) -- node {yes}(chromeassign);
\path [line] (chrome) -- node {no}(discard);

\node [block, text width=13em, below=1cm of edgeassign] (consent) {\textbf{Consent Form}};

\node [block, text width=13em, below=.45cm of consent] (knowdns) {\textbf{Questions about prior knowledge of DNS}};

\node [block, text width=20em, below=.45cm of knowdns] (interfaceinitial) {\textbf{Randomly assigned and anonymized encrypted DNS interface}};

\node[block, text width=12em, below left=.65cm and -3.25cm of interfaceinitial] (initialedge) {\textbf{Shown Edge interface}\\n=48};

\node[block, text width=12em, right=.5cm of initialedge] (initialff) {\textbf{Shown Firefox interface}\\n=45};

\node[block, text width=12em, right=.5cm of initialff] (initialopera) {\textbf{Shown Opera interface}\\n=40};

\node[block, text width=15em, left=.5cm of initialedge] (initialchrome) {\textbf{Shown Brave/Chrome interface}\\n=51};

\node [block, text width=20em, below=2.3cm of interfaceinitial] (initialdecision) {\textbf{Asked about their reasoning for\\selecting encrypted DNS settings options}};

\node [block, text width=28em, below=.5cm of initialdecision] (privatedns) {\textbf{Shown Private DNS interface and asked questions about their decision making process for encrypted DNS setting }};

\node [block, text width=25em, below=.5cm of privatedns] (explanation) {\textbf{Shown an explanation of DNS/encrypted DNS and asked questions about preferences}};

\node [block, text width=28em, below=.5cm of explanation] (seconddecision) {\textbf{Shown the same encrypted DNS interface as\\earlier and asked why they selected the settings they did}};

\node [block, text width=20em, below=.5cm of seconddecision] (gereraldecision) {\textbf{Asked questions about their decision making process for encrypted DNS setting }};

\node [block, text width=20em, below=.5cm of gereraldecision] (resolverknowledge) {\textbf{Asked about knowledge and trust of different stake holders in the DNS}};

\node [block, text width=20em, below=.5cm of resolverknowledge] (ownbrowser) {\textbf{Asked to check the encrypted DNS settings in their own browser}};

\node[block, text width=5em, fill=green!20, below=.75cm of ownbrowser] (ownedge) {\textbf{Check Edge settings}\\n=34};

\node[block, text width=10em, fill=yellow!20, right=.5cm of ownedge] (ownff) {\textbf{Check Firefox settings}\\n=47};

\node[block, text width=10em, fill=red!20, right=.5cm of ownff] (ownopera) {\textbf{Check Opera settings}\\n=4};

\node[block, text width=10em, fill=violet!20, left=.5cm of ownedge] (ownchrome) {\textbf{Check Chrome settings}\\n=71};

\node[block, text width=10em, fill=orange!20, left=.5cm of ownchrome] (ownbrave) {\textbf{Check Brave settings}\\n=26};

\node [block, text width=20em, below=.75cm of ownedge] (ownchange) {\textbf{Asked to report their current settings and if they remembered modifying them}};

\node [block, text width=13em, below=.5cm of ownchange] (demographics) {\textbf{Demographics}};


\path [line] (consent) -- (knowdns);
\path [line] (knowdns) -- (interfaceinitial);

\path [line] (interfaceinitial) -- (initialedge);
\path [line] (interfaceinitial) -- (initialchrome);
\path [line] (interfaceinitial) -- (initialff);
\path [line] (interfaceinitial) -- (initialopera);

\path [line] (initialedge) -- (initialdecision);
\path [line] (initialchrome) -- (initialdecision);
\path [line] (initialff) -- (initialdecision);
\path [line] (initialopera) -- (initialdecision);

\path [line] (initialdecision) -- (privatedns);
\path [line] (privatedns) -- (explanation);
\path [line] (explanation) -- (seconddecision);
\path [line] (seconddecision) -- (gereraldecision);
\path [line] (gereraldecision) -- (resolverknowledge);

\path [line] (resolverknowledge) -- (ownbrowser);

\path [line] (ownbrowser) -- (ownedge);
\path [line] (ownbrowser) -- (ownchrome);
\path [line] (ownbrowser) -- (ownbrave);
\path [line] (ownbrowser) -- (ownff);
\path [line] (ownbrowser) -- (ownopera);

\path [line] (ownedge) -- (ownchange);
\path [line] (ownchrome) -- (ownchange);
\path [line] (ownbrave) -- (ownchange);
\path [line] (ownff) -- (ownchange);
\path [line] (ownopera) -- (ownchange);

\path [line] (ownchange) -- (demographics);

\end{tikzpicture} } 
    \caption{Overview of Survey Structure.}%
    \label{fig:method_diagram}
\end{figure}
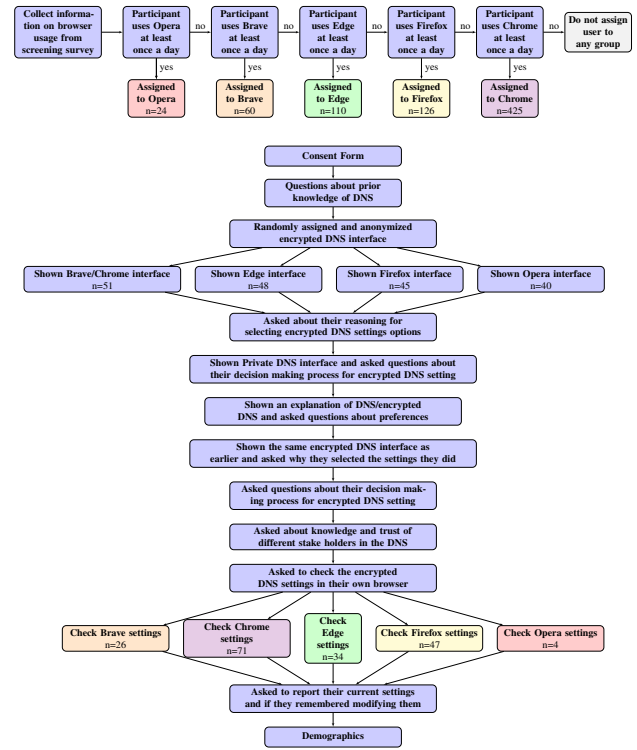

\begin{table}[t]
    \centering
    \begin{tabular}{|l|r|r|}
        \hline
        \textbf{Browser} & \textbf{\# in pool} & \textbf{ \# took survey} \\ \hline
        Edge & 110 & 50 \\ 
        Chrome & 424 & 50 \\ 
        Brave & 56 & 34 \\ 
        Firefox & 126 & 50 \\ 
        Opera & 23 & 5 \\ \hline
        \textbf{Total} & \textbf{739} & \textbf{189} \\ \hline
    \end{tabular}
    \caption{Eligible survey participants and assignments to subgroups.}
    \label{tab:survey-participation}
\end{table}
\paragraph{Overview: Two-Phase Survey.} 
Before starting either of the surveys, respondents were asked to read a
consent form and agree to participate in the study.  If a participant was not
eligible to participate (e.g., only having access to browsers not included in
our study), the participant was immediately redirected to a
survey termination page.  In the initial survey, we asked 800 participants
about their use of different popular web browsers as well as filler questions
about their mobile and Internet service providers.  
Based on their responses
to these questions, we assigned participants to subgroups determined by the
browsers they reported using at least once a week.
Table~\ref{tab:survey-participation}
shows the distribution of users who were assigned to each sub-group.  We included users who
reported using multiple browsers at least once a week in the grouping for the
least commonly used browser they mentioned, to balance out sample sizes across
browsers, and to ensure that we recruited participants who used a variety of
major browsers that offer encrypted DNS.  We excluded participants that failed
attention checks, as indicated by the differing number of participants at
different phases of Figure~\ref{fig:method_diagram}.  

\paragraph{Participant Assignment.}
After assigning participants to sub-groups according each browser type, we invited up
to 50 participants from each group to participate in a second, longer survey.
For some of the less-commonly used browsers (Brave and Opera), we were only
able to recruit a lower number of participants.  In the second survey,
following the consent form, we asked participants if they had heard of the DNS
prior to this study, and if they knew what DNS did.  To discourage dishonest
responses, we informed participants that they did not need to have prior
knowledge of the DNS to continue with the survey.  After this step, we
randomly assigned participants to an anonymized version of actual DNS settings
that can be found in browsers.  Because we only had a small number of
participants who used browsers with smaller market shares, our intent with
randomization was to achieve a better sample of responses for interfaces with
less-popular browsers, without biasing those responses based on the user's
current browser.
Rather than simply
showing them a screenshot of their assigned interface or just giving them a
multiple choice question with the different settings options, we embedded
interactive versions of the browser directly in the survey so that
participants could interact with them in the same way they would in their own
browser.  

\paragraph{Data Collection.}
We recorded information about how participants interacted with the interface
they were shown. We logged not only the final options that they selected, but
also how they interacted with the interface before making a final selection to
gain better information about whether a user may have been confused by
settings options or interfaces.
Before being shown the interface, each user was asked to select the setting 
that they would choose if they encountered it in their own browser.  These
interfaces were based on Chrome, Brave, Edge, Firefox, and Opera.  We removed
any reference to the browser itself from the interface to reduce potential
reputation bias, but we opted to leave the actual company names for any
recommended resolvers as this best simulates a real browser interface. 
We merged the Chrome and Brave interfaces because the interfaces used the same
terminology and layout; we showed users all of the resolvers that are available in
both browser interfaces.
After interacting with the settings screen, we asked all participants about
their rationale for their choosing different options, as well as their
understanding and perceptions of the different options. 
We also showed participants an interactive version of the Private
DNS settings on Android and asked about the settings that they would choose.

After collecting participants' responses, we then described the DNS and
checked the participants' understanding through a multiple-choice question
about the basic functionality of DNS.  
To ensure that our descriptions of DNS and encrypted DNS were understandable, we
performed an informal pilot with six people from various backgrounds and
iteratively edited the description. 
To ensure that our analysis of this
part of the survey was based on responses from users who had a basic
understanding of DNS, we excluded participants who did not correctly answer
the multiple-choice question from the analysis. 
For participants who met the inclusion criteria, we then explained, in more
detail, how the specific settings work and asked for their thoughts.

We then showed participants the interactive encrypted DNS settings screen that
they saw earlier in the survey and asked them once again to decide what
options they would choose and why.  This allowed us to understand how
respondents changed their settings after gaining a high-level understanding of
what encrypted DNS does and the benefits it provides. We subsequently asked
participants if they had heard of encrypted DNS before the survey.  (We
decided to ask this question after participants chose their setting in the
anonymized interfaces because we did not want to bias them while they looked
at the simulated settings interfaces that referred to encrypted DNS through
different names.)  We also briefly asked participants about their preferences
between different ways of distributing DNS queries and other security
trade-offs to better understand how participants viewed tradeoffs between
usability and privacy associated with encrypted DNS.

The next section of the survey asked participants about their knowledge
and trust in different DNS resolvers and other parties that are relevant to
DNS, such as Internet Service Providers (ISPs) and browsers.  Subsequently, we
asked participants to report on the current DoH settings in a browser that
they regularly use. We also asked about their experience with these settings.
The browser that we showed to each participant depended on the
respective response to the screening survey: If, for some reason, a participant did not
have access to the browser subgroup that we assigned them to, we reassigned
the participant to another browser that they did have access to.  We then
asked participants about the operating system of their primary mobile phone.
If a participant reported using a phone with an operating system that supported
DoT, we asked the participant to check their DNS settings and answer questions
about their use of those settings.
We concluded the survey with basic demographics and an optional question for
participants to leave feedback.

\paragraph{Data Analysis.}
We collected both quantitative and qualitative data.
Qualitative coding was performed on all free-response questions that participants were shown.
For each question, one member of the research team read through all of the responses to create a codebook and used it to assign codes to each response.
A second member of the research team coded the same responses using the codebook created by the primary coder.  
All disagreements in codes were resolved through discussion between the primary and secondary coders.
As an attention check, we excluded participants that incorrectly answered the question about the
function of DNS after our explanation as discussed in Section~\ref{sec:method}
or answered free response questions with responses that were unrelated to the
prompts. 
The statistics reported in this paper are statistically significant, based on
the results of a paired t-test with a significance level of 0.05.

\subsection{Recruitment} 

We recruited participants via Prolific, where
we required that they were at
least 18 years old, live in the US, and have completed at least 100 surveys before the
first survey with at least a 95\% approval rating, all common techniques to increase the likelihood
of quality crowdsourced responses.  The recruitment text of
the survey was phrased as a set of surveys where individuals would be asked
about network settings and did not mention encrypted DNS or security to avoid
self-selection bias.  Participants were also asked to complete the survey on a
desktop or laptop computer to avoid formatting issues.  The first part of the
survey was designed to take approximately four minutes while the second part
was designed to take approximately 20 minutes.  Participants were paid \$0.50
and \$3.30 for completing the first and second parts, respectively.  Payments
were made within 72 hours of the participants completing the survey through
the Prolific platform.  We only compensated participants
who completed the full survey. Table~\ref{tab:demographics} summarizes the
demographics of our participants. Figure~\ref{fig:method_diagram} illustrates the flow of the survey.

\begin{table}[h!]
    \centering
    \small
    \begin{tabular}{lr|lr}
    \hline
    \textbf{Gender} & \textbf{\#} & \textbf{CS or CE or IT Background} & \textbf{\#} \\
    \hline
    Female & 75 & Yes & 38 \\
    Male & 109 & No  & 140 \\
    Non-binary & 0 & Prefer not to answer & 6 \\
    Prefer to self describe & 0 \\
    Prefer not to answer & 0  \\
    \hline

    \textbf{Age} & \textbf{\#} & \textbf{Education} & \textbf{\#} \\
    \hline
    18 - 24 & 18 & Less than high school & 2 \\
    25 - 34 & 59 & High school graduate & 20 \\
    35 - 44 & 47 & Some college & 31 \\
    45 - 54 & 32 & 2 year degree & 17 \\
    55 - 64 & 15 & 4 year degree & 74 \\
    65 - 74 & 11 & Professional degree & 32 \\
    75 - 84 & 2 & Doctorate  & 6  \\ 
    Prefer not to answer & 0 & Prefer not to answer & 2 \\
    \hline

    \label{table:demo}
    \end{tabular}
    \caption{Participant demographics.}\label{tab:demographics}
\end{table}

\subsection{Limitations}\label{sec:Limitations}\label{sec:method:limitations}

This study design has several limitations. First, the encrypted DNS interfaces
were shown to participants in the context of a survey, rather than in the
context of other settings in their browser's settings menu.  Showing users
these menu options in the specific context of a study and survey on encrypted
DNS might in some cases affect the nature of responses, given that users were
only presented the part of the configuration menus specifically associated
with encrypted DNS.  Informing participants that their choice of encrypted DNS
setting would not affect the settings of their actual browser may also have
influenced the options that they chose in the survey, since users knew that
their choices in the survey would ultimately not have any practical effect on
their own user experience, or their privacy. The survey-based setup also
prevented users from experimenting with different settings. In an actual
browser, a user might experiment with several settings to see their impact,
which was not possible in this survey.

Some of our survey questions, such as those concerning the operation and
functioning of the DNS, asked participants to take their best guess, rather
than asking them to specify an option such as ``I don't know''.  This was
done to avoid participants from guessing the correct answer, which would have 
introduced bias into the survey.  However, this also means that we cannot
determine how many participants did not know the answer to a question (i.e.,
it is impossible for these questions to distinguish a correct random guess
from a user who actually knew the correct answer). Providing ``I don't know''
options has also been shown to cause participants to disengage from
surveys~\cite{qualtrics-dont-know}. It also might have made sense
to provide a ``no basis to judge'' option for the question concerning trust in
resolvers; we did ask users about their familiarity with each of these
providers and did perform a separate analysis based on this subset, with no
substantial differences in trends.

The participant sample also has some limitations, as the sample demographic
(Prolific users) may not directly correspond to a population sample for the
relevant target population (all browser users). Our sample of participants was
skewed towards male participants. 
The results of this survey may thus not necessarily generalize to a broader
population. Nevertheless, these survey results are still useful because our
goal in this work is not to make general statistical claims about the broader
population, but rather to gain insight into how users make choices with regard
to encrypted DNS settings. In this case, a sample of the general population is
likely to shed light on similar issues and insights as they pertain to
encrypted DNS. It may not be advisable to cite percentages of respondents in
this paper as they might pertain to the general population, but we do expect
that the trends that appear in this study would also be {\em present} in the
general population and thus, the issues that come to light in this study are
some that designers of encrypted DNS interfaces should consider.

Limited attention is always a potential limitation.  We mitigate this concern
by removing participants that did not understand the description of DNS or
gave answers to free-response questions that were consistently unrelated to
the questions, as a form of attention check.

Finally, our survey design potentially introduces some ordering effects
when attempting to understand the effects of prompting and education on how
users make choices about encrypted DNS settings (e.g., survey respondents
might feel as though they need to change their answers because they are being
shown the same question once again). An alternative strategy would have been
to conduct a randomized controlled trial, with a subset of participants given
the explanation at the beginning of the survey instead of in the middle. We
ultimately did not take this approach due to complications of implementing a
randomized controlled trial (RCT),
such as full blinding, smaller sample sizes due to stratification; thus, our
claims about the effectiveness of user education concerning encrypted DNS
should likely be compared against a future RCT-based study.  Nonetheless,
although our results should not be used to draw causal relationships between
education and informed choice, the effects of user prompting about encrypted
DNS settings (as might be done in real-world interfaces) are nonetheless
valid.

\subsection{Ethics}
\label{sec:method:ethics} 

This study was approved by our university's Institutional Review Board (IRB)
and was designed in accordance with the ethical principles outlined 
in the Belmont Report:
(1) respect for humans; (2) beneficence (risk vs. benefit); (3)
justice (beneficiaries versus those who bear the risks). With regard to
respect for humans, as the reviewers point out, tracking is naturally a
concern. 

Respect for humans was considered as part of the consent process and survey
design: Due to the short nature of the screening and main surveys, participant
fatigue was not expected or reported.  Before taking either the screening or
main surveys, participants provided their consent to participate via a form,
which informed them about the structure of the survey and their rights as a
participant.  When participants were shown the simulated encrypted DNS
settings interface, they were informed that their choice of setting would not
affect the setting in their own browser.  Participants potentially garnered
some benefit from taking this survey by having the opportunity to learn more
about the potential risks and benefits of encrypted DNS settings which might
enable them to make more informed DNS settings choices in the future.  From
the perspective of justice, the benefits of this survey are likely to benefit
the same population as that of the respondents: Internet users who depend on
web browsers and mobile operating systems (and the underlying DNS) to access
websites and Internet services.  Over the course of the survey, we did not
collect any personally identifiable information (PII) beyond general
demographic information.  

We note the low-risk nature of the data that we collected, specifically: the
(a) timing of clicks and (b) the state of the interface within the embedded
HTML.  We used this timing data as both a sanity check to see that the HTML
was functioning properly, as a redundant check to validate that users’ final
decisions were properly recorded, and for specific analysis (e.g., if users
even saw the resolvers listed in drop-down menus).  Note that all data
collected was exclusively within the context of the survey itself; we do not
collect any information about the user’s interactions with their browser
outside of the HTML interface embedded directly in the survey, or even
interactions with other parts of the survey.  The timing data contains the
number of milliseconds since the page loaded and the interaction the user took
at that time (e.g., toggled encrypted DNS off, opened drop-down menu of DNS
resolvers, the incorrect URL format warning shown, etc.). We implemented these
interfaces, adding code to observe user clicks on specific elements of
interest in the custom HTML and a log of the resulting effects from the code
(e.g., the drop-down menu closed because they clicked on the “my current
service provider” option).

Due to both the minimal nature of the data collected (i.e., timing data only)
and the context in which it was collected (i.e., exclusively within the
context of the survey form), we reasoned that there was minimal risk of harm
(the IRB concurred). The users consented to participating at the beginning of
the survey, but disclosing that we were timing certain interactions could have
invalidated our results, because it would have impacted how they interacted
with the interfaces or survey at large (e.g., knowledge about being timed or
otherwise under time pressure might affect their process, such as rushing to
make a decision). Given the minimal risk, we applied the Belmont Report
principle of beneficence, reasoning that the benefits of not disclosing this
part of the process to users outweighed the risks of potentially invalidating
the results. 

\section{Results}\label{sec:results}

In this section, we present the results from our survey and highlight
prevailing themes that emerged during our analysis.

\begin{figure}[t!]
  \includegraphics[width=\columnwidth]{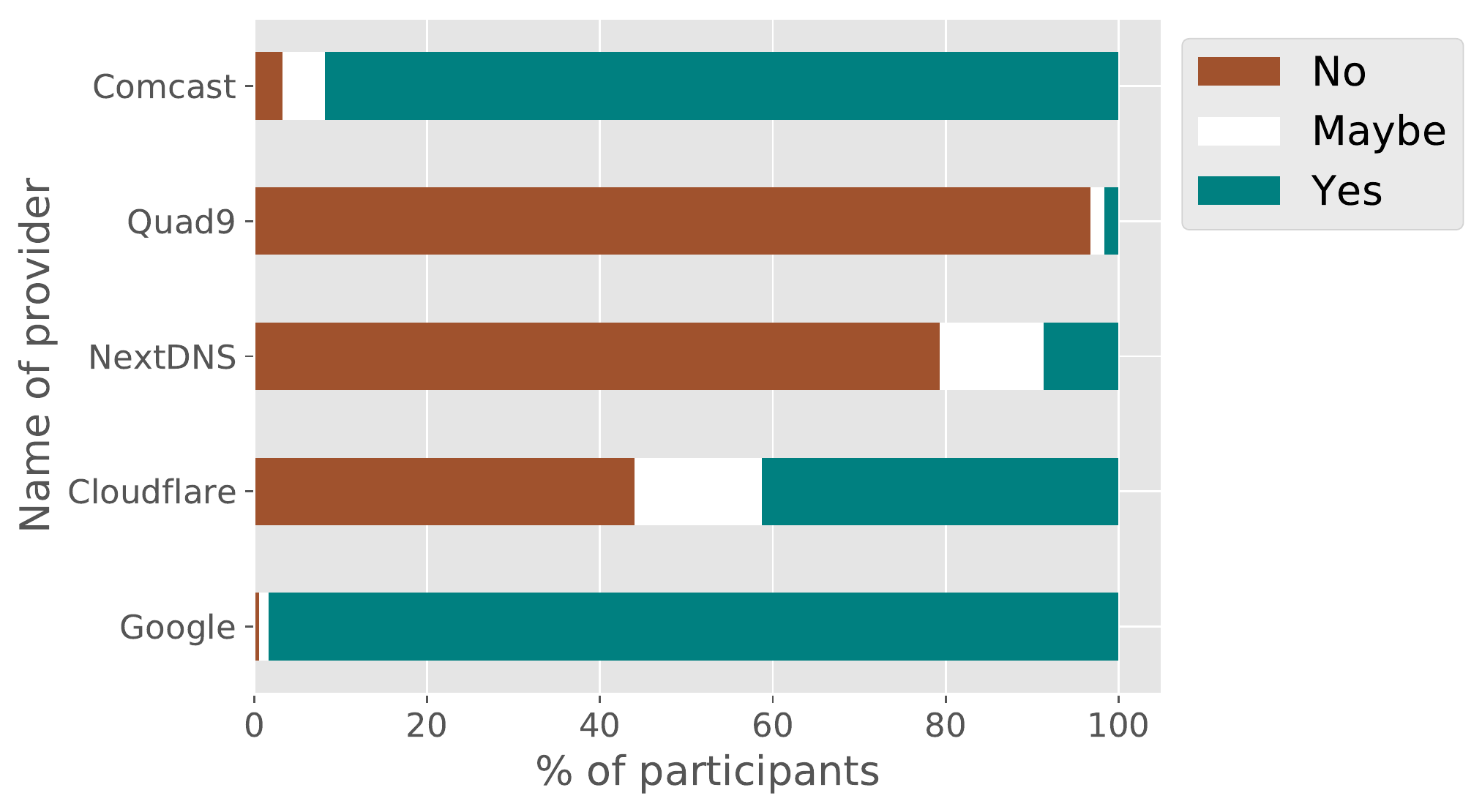}
  \caption{Percentage of participants who had heard of different stakeholders
    in the encrypted DNS ecosystem.}
  \label{fig:heard_of_provider}
\end{figure}

\subsection{Do users know about and trust encrypted DNS and its stakeholders?}\label{sec:awareness}

\begin{tcolorbox}[width=\linewidth,center]
    Many participants had heard of DNS prior to the survey, yet fewer had heard of encrypted DNS.
    Of the participants that had heard of DNS, most did not know what it did,
    and many who thought they knew what it did were unable to correctly describe its primary function.
    
    Participants were largely ambivalent about their trust in companies that
    are not well-known in contexts other than DNS, while they had stronger
    opinions on their trust in other well-known companies, as well as other
    stakeholders in the DNS ecosystem with whom they had preexisting relationships (e.g., their primary browser or ISP).
\end{tcolorbox}
\noindent
Participants frequently reported having heard of DNS, but often did not know
what it did or had incorrect assumptions about its function.  73\% of all
participants reported having heard of DNS prior to the survey, with 36\% of
the participants believing that they knew the functionality that DNS provides.
Although participants could overstate their knowledge of DNS, we attempted to
mitigate this possibility by asking the question at the beginning of the
survey and clearly stating that individuals would be able to participate in
the survey regardless of whether they had any knowledge of the DNS.
The majority of participants who self-reported having a background in computer science, computer engineering, or information technology stated that they had heard of DNS and knew what it does. 
Over three times as many participants that did not report having a background in these areas reported having heard of DNS but not knowing what it does or having never heard of DNS than reported that they knew what DNS does.

When participants who claimed to know the purpose of DNS were asked to
describe it in their own words, fewer than half mentioned the DNS as being
responsible for translating domain names to IP addresses.  The other
participants had varying misconceptions of its purpose, including identifying
computers on a network and connecting a device to the Internet. 
One participant (P98) stated that DNS was ``denial of service. blocks access to a website. prevents connection.''
Another (P95) explained, ``It is a device naming system that connects devices to the Internet with unique identifiers.''
Thus, fewer than 18\% of participants could describe what DNS does prior to the survey.
Of the participants who reported having heard of DNS, only 59.9\% had heard of
encrypted DNS.

Individuals who had heard of DNS without necessarily fully understanding it
may have encountered the term tangentially in other contexts not specific to
encrypted DNS.  Examples of when users have encountered DNS in the popular
press may include: (1)~when problems with DNS cause Internet outages and make
popular websites and services inaccessible~\cite{nytimes-outage}; or (2)~when widespread
vulnerabilities are disclosed or a major security event
occurs~\cite{nytimes-mirai}.  In these
cases, people may be exposed to the terminology and a brief description of the
DNS, yet we suspect they are likely focused on how
DNS affects them, rather than internalizing technical details.

\begin{figure}[t!]
    \centering
    \hspace*{-0.2in}
  \includegraphics[width=1.1\columnwidth]{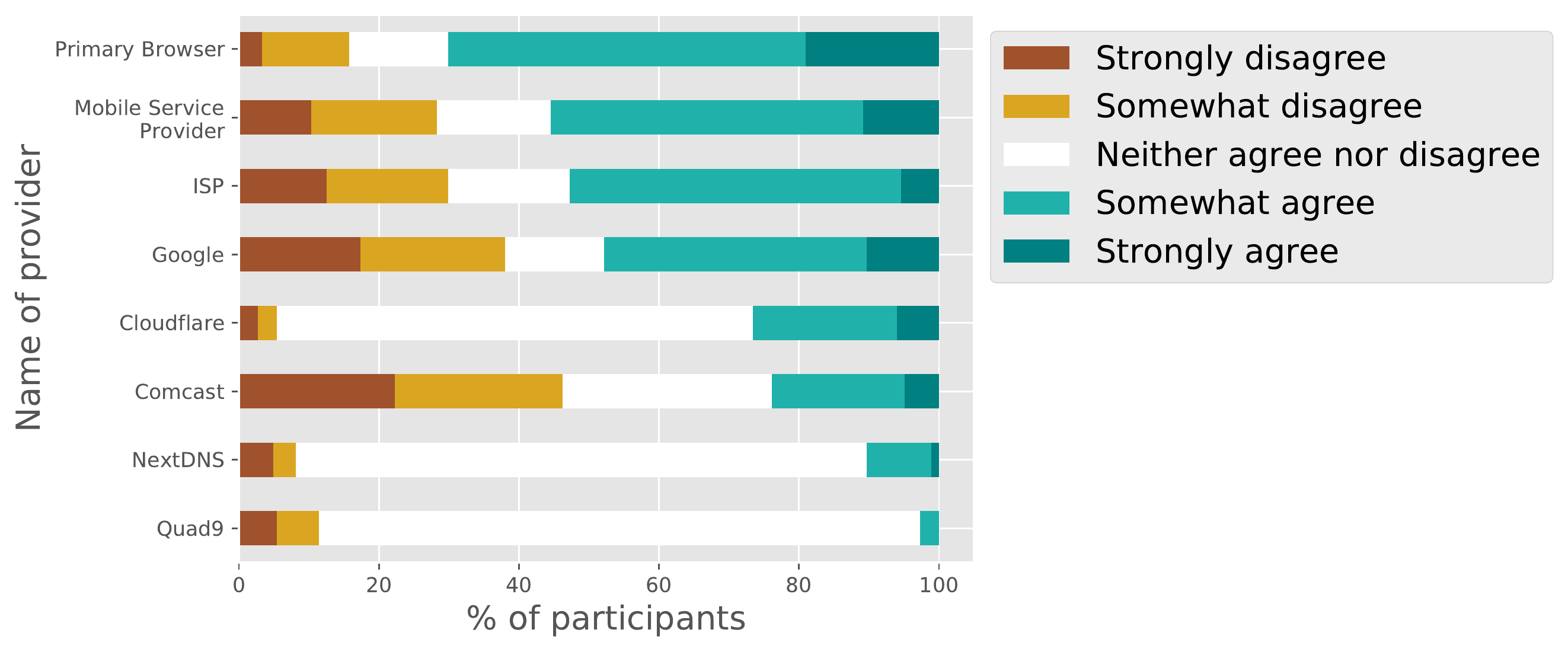}
  \caption{Percentage of participants who trust their encrypted DNS provider.}
  \label{fig:trust_provider}
\end{figure}

As Figure~\ref{fig:heard_of_provider} demonstrates, most participants had
heard of Google and Comcast but were less familiar with Cloudflare, NextDNS, and Quad 9. 
Further, Figure~\ref{fig:trust_provider} shows that participants
were ambivalent about their trust in NextDNS and Quad 9. Both of these factors may
be due to participants' lack of knowledge about these providers.
Participants were more divided on if they trusted the more well-known
companies, Google and Comcast, that also operate recursive resolvers. More participants reported that they trusted their primary browser, mobile service provider, and Internet service provider, which may
be related to their familiarity and previous interactions with them. 

Because so few participants knew what the DNS was or what DNS queries do, it
is unreasonable to expect users to understand the implications that their
choices of setting could have on their security and general browsing
experience.  Thus, even a small description or a link to a more in-depth description
of DNS and encrypted DNS could be useful when presenting these settings to
users.

\subsection{What encrypted DNS settings do users have enabled in their
browsers and mobile devices?}\label{sec:settings}

\begin{tcolorbox}[width=\linewidth,center]
    Most participants had the default settings enabled in their browsers,
    which typically meant that participants' browsers were opportunistically
    encrypting their DNS queries. The only deviations from the defaults were
    to disable the setting, or to use the Cloudflare or Google resolvers. In
    the mobile devices, to our surprise, there was a more even split between
    users that had the default ``Automatic'' (opportunistic) setting and the
    setting entirely disabled.
\end{tcolorbox}

\begin{table}[]
\resizebox{\linewidth}{!}{
\begin{tabular}{l|rl|ccccc}
Browser & \multicolumn{2}{c|}{\shortstack{Successfully\\reached setting\\(\% of participants)}} & \shortstack{With Current\\Service Provider} & Google & Cloudflare & Disabled & Unsure\\
\hline
Brave   & 23 & (88.5\%) & \cellcolor[gray]{0.8}\textbf{21}& 0 & 1 & 1 & 0\\
Chrome  & 58 & (81.7\%) & \cellcolor[gray]{0.8}\textbf{46} & 2 & 0 & 10 & 0\\
Edge    &  24 & (70.6\%) & \cellcolor[gray]{0.8}\textbf{18} & 0 & 0 & 5 & 1 \\
Firefox & 43 & (91.5\%) & N/A & N/A & N/A & N/A & N/A\\
Opera   &  3& (75.0\%) & \textbf{-} & 2 & 1 & \cellcolor[gray]{0.8}\textbf{0} & 0\\
\hline
\multicolumn{2}{l}{{\small\cellcolor[gray]{0.8} \textbf{default setting}}}\\
\end{tabular}}
\caption{Percentage of participants who reached the encrypted setting page and their subsequent resolver selections.}
\label{tab:enc_page}
\end{table}

Although most participants were able to access the encrypted DNS settings
menu, most still reported that they had the default options enabled, as Table~\ref{tab:enc_page} demonstrates. 
``With Current Service Provider'' refers to the resolver that a user's device would use by default, indicating no changes made by the user. 
108 participants reported that they could access one of the Brave, Chrome, Edge,
and Opera interfaces.\footnote{Due to an error in the survey, we do not have
the actual settings for individuals that looked at their settings in the
Firefox browser.} 
One participant was unsure if the setting was enabled.
Although the survey provided written instructions and a video describing how
to access the encrypted DNS settings in each browser, it is possible that the
participants who could not access the encrypted DNS settings menu were using
older versions of the browser that did not provide support for these settings.
Additionally, participants may have simply
been unable to find the setting due to the complexity of browser settings
menus.  Of the remaining 107 participants, 85 (79.4\%) had the default
settings for their browser selected. 

Only six participants had encrypted DNS configured in such a way that all of
their queries went to a single resolver. All of those individuals were using
either Cloudflare's resolver or Google Public DNS.  No participants reported
inputting a custom DNS resolver in their browser.  This observation
underscores not only the importance of good defaults, but also the careful
selection of the suggested resolvers that are shown to users because those are
more likely to be selected than a custom DNS resolver. Many of the participants that had a setting other than the default reported not remembering changing the setting.
This observation could result in part from circumstances, such as not having the most updated browser or operating system,
or perhaps having had settings modified by another user on a shared machine.

In the case of mobile devices, 109 participants reported having a phone that ran an Android operating system and 83 reported being able to navigate to the ``Private DNS'' settings page.
Of these participants, 48 (57.8\%) had the ``Automatic'' option selected, 34 (41.0\%) had ``Off'' selected, and only one participant had ``Private DNS Provider hostname'' selected with the correct URL to send their queries to Cloudflare.
``Automatic'' is the current default, yet none of the participants that had the ``Off'' setting selected reported remembering changing this setting in the past.
This suggests that perhaps the default of ``Automatic'' was somehow not applied to them, so while their phone supported the opportunistic mode, they were not taking advantage of the potential benefits of the setting.
Participants who were unable to reach the Private DNS settings page may have
had older operating systems installed on their phone, and thus might not be able to access the page.

\subsection{When shown encrypted DNS settings for different browsers, which settings do users select, and why?}
\label{sec:selection}

\begin{tcolorbox}[width=\linewidth,center]
    Most participants chose the default settings in the interface
    shown to them. There were variations in the settings that users chose
    based on which interface they were shown. For example, no participants
    that saw a Chromium-based interface disabled the setting. No participants correctly entered a custom
    trusted recursive DNS resolver. 
\end{tcolorbox}

\begin{table*}[t]
    \centering
    \large
    \resizebox{\linewidth}{!}{
    \begin{tabular}{l|r|r|r|rrrrrr|r}
    {\normalsize\textbf{Browser}}& {\normalsize\textbf{\#}} & {\small\textbf{Off}} & {\small\textbf{\shortstack{With Current\\Service Provider}}} & {\small\textbf{Cloudflare}} & {\small\textbf{\shortstack{Google\\(Public DNS)}}} & {\small\textbf{Quad9}} & {\small\textbf{NextDNS}} & {\small\textbf{\shortstack{CleanBrowsing\\(Family Filter)}}} & {\small\textbf{OpenDNS}} & {\small\textbf{Custom}}\\
    \hline
    {\normalsize\textbf{Chrome/Brave}} & 51    & 0 &\cellcolor[gray]{0.8}\textbf{37} & 6 & 2 & 0 & 0 & 0 & 1 & 3 \\
    {\normalsize\textbf{Edge}} & 48    & 0 & \cellcolor[gray]{0.8}\textbf{41} & 3 & 3 & 0 & 0 & 1 & 1 & 0 \\
    {\normalsize\textbf{Firefox}} & 45   & 6 & - & \cellcolor[gray]{0.8}\textbf{34} & - & - & 3 & - & - & 2 \\
    
    {\normalsize\textbf{Opera}} & 40   & \cellcolor[gray]{0.8}\textbf{20} & - & 17$^*$ & 3 & - & - & - & - & 0 \\
    \hline
    {\normalsize\textbf{Android}} & 184 & 12 & \cellcolor[gray]{0.8}\textbf{158} & - & - & - & - & - & - & 18\\
    \hline
    
    {\small\cellcolor[gray]{0.8} \textbf{default setting}} & 
    \multicolumn{6}{l}{{\small - indicates that the option is not available for that browser}}\\ 
    \multicolumn{11}{l}{{\small* The Opera interface offers three version of the Cloudflare resolver: the default (11), No Malware (2), and No Malware or Adult Content (4)}} \\
    \end{tabular}}
    \caption{Users' choice of encrypted DNS setting in the anonymized browser
    interfaces with no addition information about DNS or encrypted DNS. With
    current service provider indicates that the resolver would default to
    the trusted resolver of their ISP.}
    \label{tab:InitialChoice}
\end{table*}
\noindent
When shown the anonymized browser interfaces for encrypted DNS, 71.7\% of participants continued to use the default settings shown to them.
Participants seemed to trust the default, stating, ``Because it is the default, so I feel it is recommended by the software developers.'' (P135).
This percentage varied by browser, with Firefox having the highest percentage
of users continuing with the default setting (75.6\%) and Opera having the
lowest, with only 50\% of participants selecting the default.
Opera is the only browser that has encrypted DNS disabled by default, which could have been a factor for many people modifying the setting. 
Firefox, on the other hand, has the fewest alternatives.
The predominant reasons participants gave for selecting the settings that they
did included: (1)~a perceived increase in security, simply because the setting
was the default; or (2)~because they just didn't know what the setting did. As one participant put it (P144), ``It was the default setting and also helps prevent low level attacks from hackers.''
Participants' choices for the ``Private DNS'' setting show a similar pattern, with 83.7\% of participants keeping the default ``Automatic'' setting.
This finding illustrates the importance of how browsers and mobile service providers configure defaults.

No participants disabled the encrypted DNS setting when shown any of the
interfaces for the Chromium-based browsers (Brave, Chrome, Edge); 
86.7\% of users shown the Firefox interface and 50\% of participants shown
the Opera interface enabled the setting.
Chromium-based browsers have a small toggle that collapses the settings
window, it is possible that discouraged people from disabling the setting.
Furthermore, the option to use their current service provider might have been
seen as a lower risk alternative since they already have a relationship with
that company: ``I chose my current service provider because I trust that it is
a great choice since I am currently using it.'' was how one user expressed the
sentiment (P31).

When participants enabled encrypted DNS, but did not go with the default
setting, they were much more likely to choose a resolver that was listed in
the setting rather than specifying a custom resolver. Cloudflare was by far
the most popular. Table~\ref{tab:InitialChoice} shows the initial choices of
the users.

Although few participants chose to specify a custom resolver in any of the
browser (2.7\%) or mobile (9.8\%) interfaces, the ones who did entered text
that would not function the way they might expect it to.  \textbf{In fact, no
participant entered a custom DNS resolver that would have resulted in their
queries being encrypted.} For example ``McAfee,'' ``www.google.com,'' and
``1.1.1.1'' were entered by participants into browser and mobile interfaces.
In the case of ``1.1.1.1'', while it may seem correct, to actually use the
Cloudflare DNS resolver through the custom resolver input field, the user
would need to have entered ``1dot1dot1dot1.cloudflare-dns.com'' on android or
``https://cloudflare-dns.com/dns-query'' for any of the browsers.  Some
interfaces do check whether the text entered will function, but it is often
possible to click away from the setting before the warnings are shown. It
would be beneficial to users if all of the interfaces, rather than only some,
would give users actionable feedback on the validity of their inputs before
they exit the page.

When asked about the advantages and disadvantages of enabling encrypted DNS,
many participants thought enabling the setting would result in general
improvements to security, although they were often unable to go into any
detail as to what those security benefits might be, stating ``HTTPS has a major advantage of being more secure.'' (P44) and ``I might have more security while using my computer'' (P65). 
Based on survey responses, participants who saw interfaces that labeled DoH as ``secure DNS''
were more likely to mention security as a potential benefit of enabling DoH
than people that saw the setting called ``DNS over HTTPS'' ($p<0.05$).
This connection between the name of the setting and the perceived impact
emphasizes the importance of this subtle difference in wording.

There was confusion among participants about how their choice of encrypted DNS
resolver might affect their browsing performance: 19.6\% of participants
mentioning that the setting could slow down Internet browsing, while 13.0\%
(not a significantly different number of participants) thought it might
improve performance. Such confusion is actually consistent with empirical
studies, which have shown that the relative performance improvement (or
degradation) of encrypted DNS actually depends quite a lot on the choice of
encrypted DNS resolver and client~\cite{hounsel2021can}.

Aside from the potential effects on speed, participants also mentioned concerns
that enabling encrypted DNS might reduce their access to websites with one
respondent stating, ``If it is security-related, it may restrict access to
certain domains.'' (P6). The desire to be able to access the Internet as they
normally do was also mentioned by several participants as the reason they
choose their respective settings. Participants overwhelmingly wanted an
explanation of the settings to assist with making the decision. For example, one participant (P47) said, ``I would have wanted to know what each setting represented in simple terms.''
Another participant (P14) said, ``I would want to know what DNS stands for and what it does, as well as any non-obvious considerations I may want to think through before enabling it.''
More specific requests included wanting definitions of different relevant terms shown on the
settings page, the pros and cons of different settings, and information on the
security benefits of each setting.

Interfaces that labeled the setting with the technical name ``DNS over HTTPS'' caused additional confusion in some participants.
\textbf{Instead of interpreting the name as meaning DNS \textit{using} the HTTPS protocol they interpreted DoH as meaning use DNS \textit{instead} of HTTPS.}
Of the participants who saw the Firefox or Opera interfaces, that label the setting in this way, 10.6\% mentioned an incorrect interpretation of the setting name as part of their reason for choosing the setting option that they did. 
There may have been more participants who misinterpreted the setting name, but they did not mention it in their reasoning for choosing their preferred setting option.
One participant (P3) who saw the Firefox interface and chose to disable DoH stated, ``I have no earthly idea what DNS is, while I at least have a passing familiarity with HTTPS.'' Another participant (P30) said ``From the little I know I believe that HTTPS is more secure than DNS'' and chose to disable the setting in the Opera interface.
While most of the participants who misinterpreted the setting name in this way ended up opting to disable the setting, that decision was not universal. 
This observation highlights the importance of avoiding technical jargon that could be easily misinterpreted by average users.

\begin{table*}[t]
    \centering
    \large
    \resizebox{\linewidth}{!}{
    \begin{tabular}{l|r|r|r|rrrrrr|r}
    \hline
    {\normalsize\textbf{Browser}}& {\normalsize\textbf{\#}} & {\small\textbf{Off}} & {\small\textbf{\shortstack{With Current\\Service Provider}}} & {\small\textbf{Cloudflare}} & {\small\textbf{\shortstack{Google\\(Public DNS)}}} & {\small\textbf{Quad9}} & {\small\textbf{NextDNS}} & {\small\textbf{\shortstack{CleanBrowsing\\(Family Filter)}}} & {\small\textbf{OpenDNS}} & {\small\textbf{Custom}}\\
    \hline
    {\normalsize\textbf{Chrome/Brave}} & 51& 1 & \cellcolor[gray]{0.8}\textbf{33} & 8 & 3 & 0 & 1 & 0 & 2 & 3\\
    {\normalsize\textbf{Edge}} & 48  & 2 & \cellcolor[gray]{0.8}\textbf{29} & 6 & 4 & 0 & 1 & 1 & 3 & 2\\
    {\normalsize\textbf{Firefox}} & 45 & 6 & - & \cellcolor[gray]{0.8}\textbf{28} & - & - & 7 & - & - & 4\\
    
    {\normalsize\textbf{Opera}} & 40 & \cellcolor[gray]{0.8}\textbf{17} & - & 19$^*$ & 4 & - & - & - & - & -\\
    \hline

    {\small\cellcolor[gray]{0.8} \textbf{default setting}} & 
    \multicolumn{6}{l}{{\small - indicates that the option is not available for that browser}}\\ 
    \multicolumn{11}{l}{{\small* The Opera interface offers three version of the Cloudflare resolver: the default (11), No Malware (5), and No Malware or Adult Content (3)}} \\
    
    \end{tabular}}
    \caption{Users' choice of encrypted DNS setting in the anonymized browser interfaces after DNS and encrypted DNS has been explained.}
    \label{tab:SecondChoice}
\end{table*}

\begin{figure*}[t!]
    \centering
  \includegraphics[width=\linewidth]{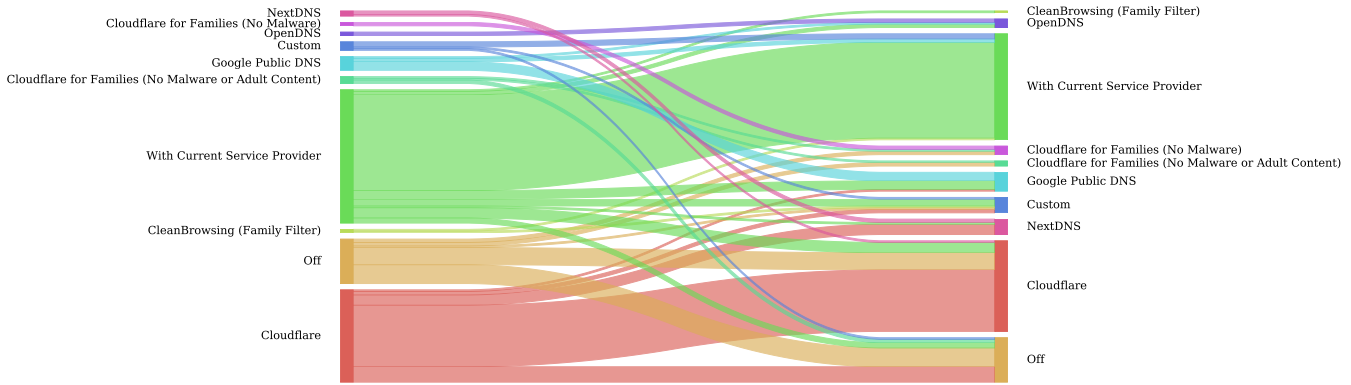}
  \caption{How users' settings choices changed from their initial choice to their decision after DNS and encrypted DNS has been explained.}
  \label{fig:sankey}
\end{figure*}

\subsection{When the technical aspects of these systems are explained to users,
how do their choices of settings change?}\label{sec:technical}

\begin{tcolorbox}[width=\linewidth,center]
    Almost 40\% of participants modified their settings in some way after
    being shown an explanation of DNS and encrypted DNS. The default settings
    remained popular, but much less so than when participants had made
    decisions
    without extra information about encrypted DNS. Although participants
    appeared to do demonstrate a better qualitative understanding of encrypted
    DNS, they still had problems understanding the differences in
    functionality or privacy guarantees between the different resolvers.
\end{tcolorbox}
\noindent
When asked to look at the same anonymous interface they were shown earlier in
the survey, after having been provided a description of what the DNS is used
for and what encrypted DNS does, 37.0\% of participants chose to modify their
settings in some way. 30 (16.3\%) participants reversed their choice: with
an equal number choosing to disable the setting that they had previously
enabled and enabling the setting that they had previously disabled.
Figure~\ref{fig:sankey} shows the relationship between the two settings that
users chose before and after receiving an explanation.

Table~\ref{tab:SecondChoice} shows all of the setting options chosen by
participants after receiving the additional information about each setting.
The default settings were still the most selected option across all of the
interfaces (59.2\%), but the number of participants who selected the
respective defaults for each browser decreased across every browser, which is
significantly different than the (71.7\%) of participants who chose the
default before they were provided with additional information (p$<$0.5).
The largest change was seen among users of the Opera browser, with 47.5\% of participants choosing to modify their setting in some way.
Participants were still more likely to choose a resolver listed in their
respective interfaces. Although more people opted to enter custom
resolver hostnames not a single participant added text that
would have functioned properly. Because the descriptions we provided to
participants did not provide instructions for how to select a resolver, but
mainly talked about the settings on a high level, such errors could be
expected and were consistent with how participants described their experience; for example, ``I understand more about DNS now, but still don't know who else I'd use if not my current service provider'' (P46).

Perceived benefits to security remained a predominant factor in many
participants' decision-making process. Unlike with their initial choice of
setting, privacy and encryption factored into participants' decision-making
processes at the end of the survey. For example, when asked why they enabled DNS-over-HTTPS, a participant (P 26) stated, ``Private browsing and a faster and secure network.'' 
Others stated ``Seems like it is the correct call for security'' (P 61) and ``Because from what I understand, the DNS gives some privacy so Internet providers don't see your online activity'' (P 94).
Because our descriptions of DNS and encrypted DNS discussed those topics, this result was expected to some degree.
Finally, prior knowledge of and trust in a company was a factor users mentioned when explaining why they chose the setting options that they did.
When asked why they chose a particular company, one participant stated, ``because I already have a relationship with them.'' (P 98).
When participants reported having heard of encrypted DNS, they were significantly more likely to enable encrypted DNS as their initial choice when they were not provided with any additional information about the setting (p$<$0.05). 
On the other hand, after everyone had been provided with information about DNS
and a brief description of encrypted DNS the differences were no longer
significant.  Note that future work could explore these effects in more detail
in a larger randomized controlled trial, as discussed in
Section~\label{sec:method:limitations}.








\section{Discussion}\label{sec:disc}

The results from previous sections highlight several general takeaways and
point to a variety of recommendations, both for designers and policymakers
(e.g., standards bodies, or regulators who may wish to standardize how various
protocol options are presented to users).

\subsection{Takeaways from Results}

Our results highlighted that, although many users have generally heard of
 DNS and that various configuration options are possible, most users do
not change their browser or mobile OS settings from the defaults, and many
users also do not understand either how DNS (or encrypted DNS) works, or the
guarantees that encrypted DNS can provide which is consistent with research that looked at other privacy settings~\cite{habib2022_okaywhatever, Liu2011FbPrivSettings, nisenoff-21-PrivateDNS}
Users generally want more information about the privacy benefits that encrypted DNS can provide, as well as information about their options in configuring it.  
These observations are
consistent with Clark's discussions of ``tussle spaces'', which have noted
that the choices that users have (and are aware of) can have significant
consequences. As such, encrypted DNS (and interface) implementations should recognize these tussles and make choices available to users in ways that allow them to make appropriate tradeoffs between privacy and performance, according to their own preferences.

Additionally, users were concerned that their attempts to customize encrypted
DNS configurations could cause basic functionality to break, resulting in
their inability to use the Internet. As it turns out, such concerns are not
unfounded. For example, specifying an custom recursive resolve with incorrect
syntax in the mobile OS configuration does result in a silent connectivity
failure, with no error message to the user concerning the nature of the
misconfiguration.

Based on these observations, we provide a set of recommendations for designers
of encrypted DNS interfaces on user-facing devices (e.g., browsers, mobile
OSes) that could allow users to make more informed choices concerning the
configuration of encrypted DNS.

\subsection{Design Recommendations}

\paragraph{Provide a basic primer on DNS function (and privacy risks).} Many
users are not aware of the functions of DNS, as well as the privacy risks
associated with the ability of a third party to observe DNS traffic.
We thus recommend that application designers find interface-agnostic ways to provide
information to users about DNS function, privacy risks, and the tradeoffs
associated with each setting. 
In some cases, it may be useful to
augment the {\em interface} itself in a way that indicates to the user which
entities and organizations can see DNS traffic for different settings, in
simple terms (e.g., ``your ISP'', ``the coffee shop's provider'', ``the web site
host'').  The exact design of such an interface
could be a ripe topic for future work, as we discuss in more detail below.


\paragraph{Provide privacy policies for the resolvers.} In principle, users
have many choices for trusted recursive resolvers, from major providers (i.e.,
Google, Cloudflare), to medium-sized operators (i.e., Quad9), to smaller
independently operated recursive resolvers. A user's choice of trusted
recursive resolver has significant implications for privacy, since the
organization operating the trusted recursive resolver sees potentially all of
the user's DNS traffic (and hence may be able to infer much about the user,
from browsing patterns to other behavior). Our results indicated that while
participants understood the setting after it was described to them, they still
struggled to understand the differences between different choices of trusted resolvers.
Because users have significantly different levels of trust for the respective operators of recursive resolvers, the privacy practices and policies of operators should be more transparent to
users to ensure that users can make more informed choices.

\paragraph{Be thoughtful about the use of technical protocol terminology,
which may not map to users' mental models.} Some technical terminology to
describe encrypted DNS protocols can be confusing to users: In particular, we
found that many users misinterpreted the phrase ``DNS-over-HTTPS'', to mean
that DNS would be used instead of HTTPS, {\em not} that HTTPS was the
transport protocol over which DNS queries and responses were transmitted. In
such cases, understanding the assumptions that users make about functionality
based on language choice can help designers choose terminology and phrasing
that better reflects the properties that a protocol provides. User studies and
focus groups may be appropriate when deploying such protocols and variants,
both now and in the future.

\paragraph{Provide users with resolver options.} The large collection of data
by one or a few mainstream resolvers raises privacy concerns.  Per the
suggestions of participants, browsers could also add more information about
the advantages and drawbacks of different choices of encrypted DNS resolvers,
which would allow them to make an educated decision about their browser
settings. 

\paragraph{Provide users with the necessary format to select a custom
resolver, and check that the user specification is correct and functional.}
Participants expressed concerns about experimenting with the settings,
fearing that they would break elements of their browsers.  Browsers could add
more apparent instructions, warnings, or guidelines to their interfaces to
provide more clarity for users.  Many survey participants also maintained the
default setting.

\subsection{Future Research}\label{sec:disc:future_work}

Future work could replicate this study on a larger scale, and across a wider
range of demographics. Because encrypted DNS is not limited to the United
States, a larger study that captures a broader cross-section of users could
deepen our understanding of user perceptions by including participants who
live in other countries.  Involving more participants, could also provide data
that may highlight broader themes, including how various attitudes and
awareness might vary according to user demographics such as age, level of
education, and geography.  Future studies could further explore  user behavior
in the context of their own browsers and mobile operating systems, rather than
in the context of a survey.  Finally, another avenue for future work could
attempt to design new interfaces that incorporate background information users
might need to make a more informed choice.

As opportunistic encrypted DNS becomes more widely adopted, default settings
and their implications will become more important. For example, this study
explored the extent to which users trust various service providers who offer
encrypted DNS; given that opportunistic encrypted DNS is becoming more
widespread, other risks, such as downgrade attacks whereby behavior reverts to
unencrypted DNS, may become more prevalent, giving rise to the need to assess
user understanding of these more subtle issues. We note that in many cases,
given current interfaces, the fallback behaviors were unclear, even to
us---suggesting the possibility for a more detailed study on encrypted DNS
fallbacks and failure modes.  Finally, future studies might evaluate the
extent to which different phrasing and explanations of settings and options
(including the privacy implications associated with different choices of
trusted recursive resolver) might ultimately affect users' attitudes and
behaviors concerning encrypted DNS settings.
 
\balance\section{Conclusion}
\label{sec:conclusion}

The increasing deployment of encrypted DNS in browsers and mobile operating
systems has significant consequences for privacy, performance, and
reliability---particularly as vendors change default settings (often without direct notification to users). Previous research has observed that encrypted DNS is a tussle space among users, Internet service providers, and content providers, because the parties who control DNS have more ability to optimize content and services and have access to potentially sensitive information about users' browsing behaviors and activities. 
Given the significant stakes of encrypted DNS deployment, users should be able to
make informed choices about how it is configured. Interfaces should make
it easy for users to be aware of how encrypted DNS is configured, as well as
how to change default settings in accordance with their preferences. Our
findings in this research confirmed that work is needed in several areas,
including: to improve
user awareness about the privacy implications of DNS, to provide users with
information to better understand the implications of how encrypted DNS is
configured, and to design setting interfaces that make these options intuitive for users to customize. Although this paper does not offer the last word
on user attitudes and awareness about encrypted DNS, our hope is that it lays
the groundwork for more research in this area, to positively affect interfaces, standardization,
and policymaking.
 
\balance
\paragraph{Acknowledgments.}
This work was funded by NSF Award SaTC-2155128 ``Understanding Practical
Deployment Considerations for Decentralized, Encrypted DNS''. Ranya Sharma and Alexandra
Nisenoff were supported by NSF Research Experiences for Undergraduates (REU)
supplement awards under NSF Award TWC-1953513 ``Towards a Science of
Censorship Resistance''. Alexandra Nisenoff is also
supported by an NSF Graduate Research Fellowship. We thank Allison Mankin, the
anonymous reviewers, and our shepherd for comments that helped improve this
paper.



\newpage
\bibliographystyle{abbrv}
\bibliography{bibliography}

\appendix
\pagebreak
\section{Appendix}\label{sec:appendix}
\subsection{Encrypted DNS Interfaces}
\floatstyle{boxed}
\restylefloat{figure}
\begin{figure}[h!]
       \centering
       \begin{subfigure}[t]{\linewidth}
             \includegraphics[width=\linewidth]{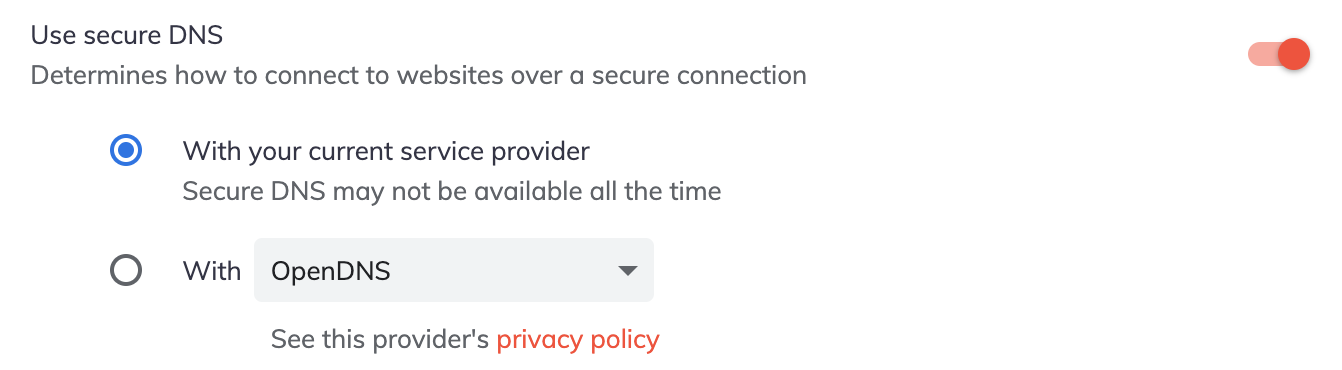}
             \caption{Brave.}
             \label{fig:brave}
       \end{subfigure}
       \begin{subfigure}[t]{\linewidth}
             \includegraphics[width=\linewidth]{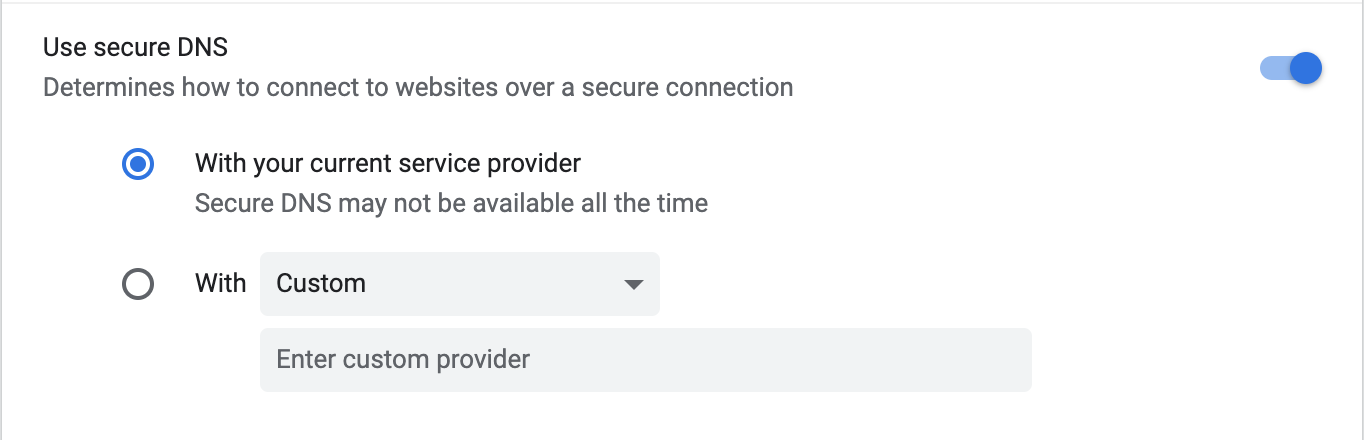}
             \caption{Chrome.}
             \label{fig:chrome}
       \end{subfigure}
       \begin{subfigure}[t]{\linewidth}
             \includegraphics[width=\linewidth]{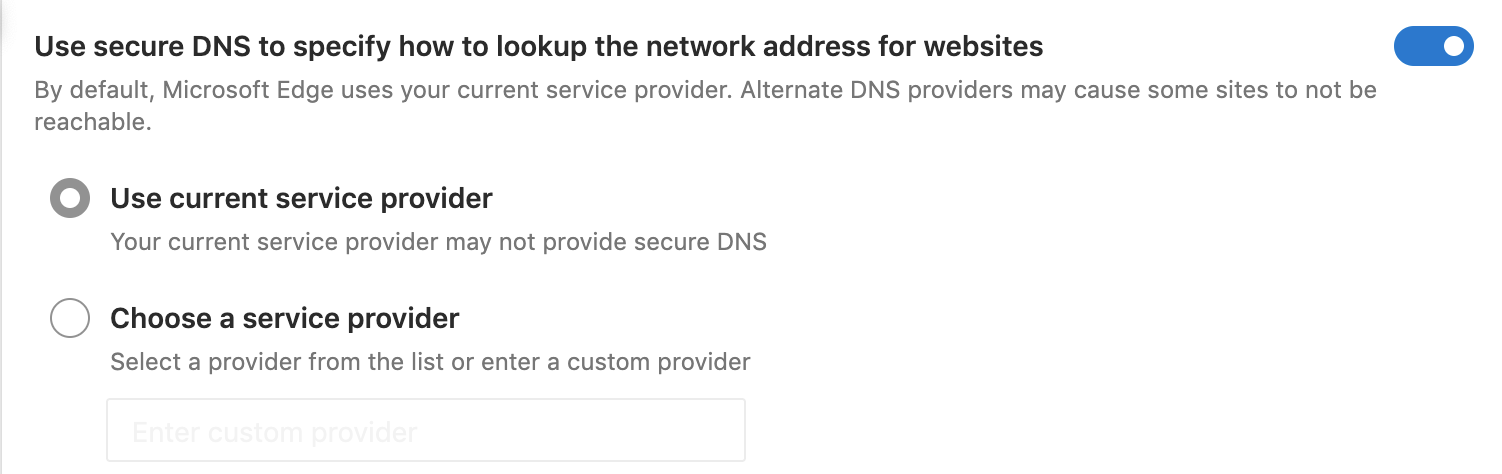}
             \caption{Edge.}
             \label{fig:edge}
       \end{subfigure}
       \begin{subfigure}[t]{\linewidth}
             \includegraphics[width=\linewidth]{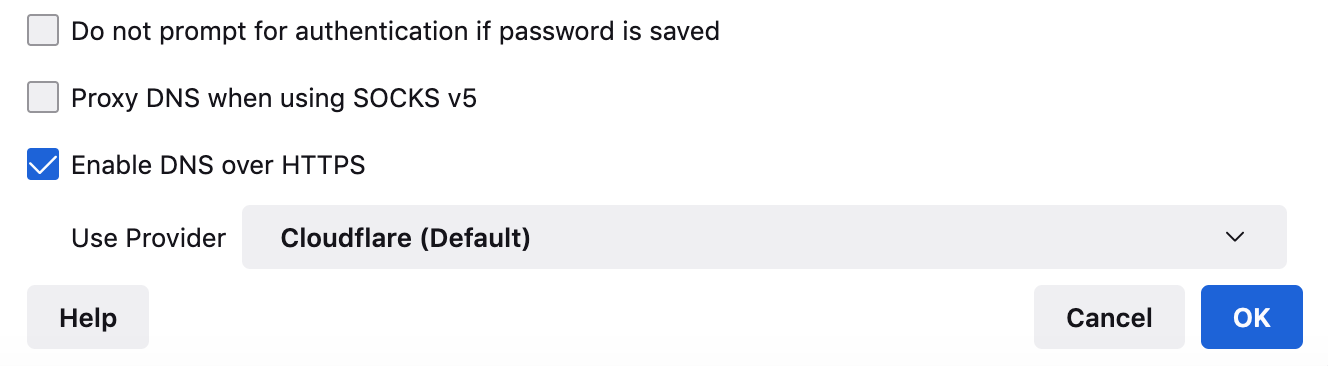}
             \caption{Firefox.}
             \label{fig:firefox}
       \end{subfigure}
       \begin{subfigure}[t]{\linewidth}
             \includegraphics[width=\linewidth]{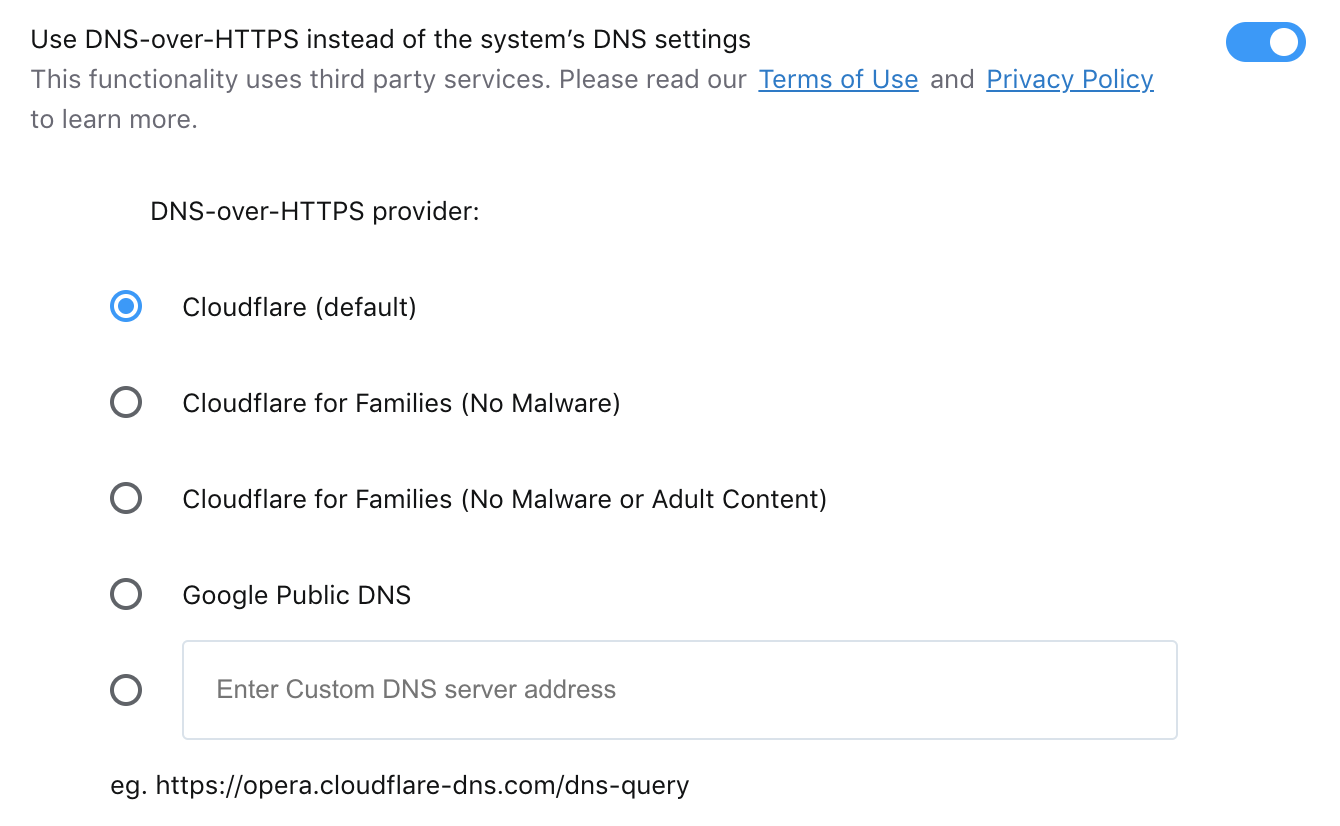}
             \caption{Opera.}
             \label{fig:opera}
       \end{subfigure}
   \caption{Encrypted DNS settings for different browsers.}
       \label{fig:enc-dns}
\end{figure}

\floatstyle{plain}
\restylefloat{figure}

\vfill
\subsection{Survey}
\subsubsection{Consent Form}

{\bf Consent Form for Participation in a Research Study}

\paragraph{Study Title:} User Expectations and Understanding of Network Settings on
Mobile Devices and Browsers 

\noindent
Principal Investigator:  \\
Other Researchers: \\
IRB Study Number: \\

\paragraph{DESCRIPTION:} We are researchers at doing a research study to better
understand what users think of network settings on mobile devices and in
browsers. In this survey you will answer questions about what you think
about settings options, your preferences among the available options,
and your opinions on what the default settings should be. Prolific
workers who are age 18+, live in the United States, and have completed
100+ studies with a 95\%+ approval rating are eligible. Participation
should take about 20 minutes.

\paragraph{RISKS and BENEFITS:} The risks to your participation in this online study
are those associated with basic computer tasks, including boredom,
fatigue, mild stress, or breach of confidentiality. The only benefit to
you is the learning experience from participating in a research study.
The benefit to society is the contribution to scientific knowledge.

\paragraph{COMPENSATION:} All participants who complete all tasks will be
compensated \$3.30 through Prolific. You will not be compensated for
partial completion of the study.

\paragraph{PLEASE NOTE:} This study contains a number of checks to make sure that
participants are finishing the tasks honestly and completely. As long as
you read the instructions and complete the tasks, your submission will
be approved. If you fail these checks, your submission will be rejected.

\paragraph{CONFIDENTIALITY:} Your Prolific ID will be used to distribute payment to
you, but will not be stored with the research data we collect from you.
Any reports and presentations about the findings from this study will
not include your name or any other information that could identify you.
In some cases, you might provide personal stories or beliefs that we
might quote or paraphrase as part of our research findings -- any
personally identifying information will be removed to protect your
privacy. We may share the data we collect in this study with other
researchers doing future studies -- if we share your data, we will not
include information that could identify you.

\paragraph{SUBJECT'S RIGHTS:} Your participation is voluntary. You may stop
participating at any time by closing the browser window or the program
to withdraw from the study. Partial data will not be analyzed.

For additional questions about this research, you may contact For
questions about your rights as a research participant, you may contact\\
Please indicate below, that you are at least 18 years old, live in the
United States, have read and understand this consent form, and agree to
participate in this online research study.

I am over 18 years old.\\
\\\hspace*{0.35in}\fullmoon~~Yes 
\\\hspace*{0.35in}\fullmoon~~No\\
 
I currently live in the United States of America. \\
\\\hspace*{0.35in}\fullmoon~~Yes 
\\\hspace*{0.35in}\fullmoon~~No\\
 
I have read and understood this consent form.\\
\\\hspace*{0.35in}\fullmoon~~Yes
\\\hspace*{0.35in}\fullmoon~~No\\
 
I agree to participate in this online research study.\\
 \\\hspace*{0.35in}\fullmoon~~Yes
 \\\hspace*{0.35in}\fullmoon~~No\\

Please enter your Prolific ID here\\

\subsubsection{Screening Survey}

How often, if at all, do you use the Chrome browser on a laptop/desktop?\\
\\\hspace*{0.35in}\fullmoon~~At least once a day
\\\hspace*{0.35in}\fullmoon~~At least once a week
\\\hspace*{0.35in}\fullmoon~~At least once a month
\\\hspace*{0.35in}\fullmoon~~Less than once a month
\\\hspace*{0.35in}\fullmoon~~I do not use this browser\\

What online activities do you regularly do in your Chrome browser? Please select all that apply.\\
\\\hspace*{0.35in}\fullmoon~~Checking the news
\\\hspace*{0.35in}\fullmoon~~Participating in online video calls or conferences
\\\hspace*{0.35in}\fullmoon~~Streaming or downloading music, radio, podcasts, etc.
\\\hspace*{0.35in}\fullmoon~~Online shopping, making reservations, or using other online consumer services
\\\hspace*{0.35in}\fullmoon~~Working remotely
\\\hspace*{0.35in}\fullmoon~~Checking the weather
\\\hspace*{0.35in}\fullmoon~~Using social media
\\\hspace*{0.35in}\fullmoon~~Text messaging or instant messaging Emailing
\\\hspace*{0.35in}\fullmoon~~Watching video online
\\\hspace*{0.35in}\fullmoon~~Using online financial services (banking, investing, paying bills, etc.)
\\\hspace*{0.35in}\fullmoon~~Other
\\\hspace*{0.35in}\fullmoon~~None of the above
\\\hspace*{0.35in}\fullmoon~~Prefer not to answer\\

What, if any, browsers other than Chrome, Firefox, Microsoft Edge, Brave, Opera, or Safari do you use?\\

What would you say is the browser you use the most?\\
\\\hspace*{0.35in}\fullmoon~~Brave
\\\hspace*{0.35in}\fullmoon~~Safari
\\\hspace*{0.35in}\fullmoon~~Opera
\\\hspace*{0.35in}\fullmoon~~Firefox
\\\hspace*{0.35in}\fullmoon~~Microsoft Edge
\\\hspace*{0.35in}\fullmoon~~Chrome
\\\hspace*{0.35in}\fullmoon~~None
\\\hspace*{0.35in}\fullmoon~~Other\\

What would you say is the browser you use the most?\\

Who is your mobile service provider (the company that you pay for phone or cellular data service on your mobile devices)?\\
\\\hspace*{0.35in}\fullmoon~~Mint Mobile
\\\hspace*{0.35in}\fullmoon~~T-Mobile
\\\hspace*{0.35in}\fullmoon~~AT\&T
\\\hspace*{0.35in}\fullmoon~~Verizon
\\\hspace*{0.35in}\fullmoon~~Visible
\\\hspace*{0.35in}\fullmoon~~Other
\\\hspace*{0.35in}\fullmoon~~I do not have a mobile service provider
\\\hspace*{0.35in}\fullmoon~~Don't know
\\\hspace*{0.35in}\fullmoon~~Prefer not to answer\\

What is your Internet Service Provider (the company that you pay for Internet access in your home)?\\
\\\hspace*{0.35in}\fullmoon~~TDS Telecom
\\\hspace*{0.35in}\fullmoon~~Consolidated Communications Cox Communications
\\\hspace*{0.35in}\fullmoon~~Verizon
\\\hspace*{0.35in}\fullmoon~~Sparklight
\\\hspace*{0.35in}\fullmoon~~CenturyLink
\\\hspace*{0.35in}\fullmoon~~Frontier Communications Comcast (Xfinity)
\\\hspace*{0.35in}\fullmoon~~Charter Communications Windstream
\\\hspace*{0.35in}\fullmoon~~Mediacom
\\\hspace*{0.35in}\fullmoon~~AT\&T
\\\hspace*{0.35in}\fullmoon~~Other
\\\hspace*{0.35in}\fullmoon~~I do not have an internet service provider
\\\hspace*{0.35in}\fullmoon~~Don't know
\\\hspace*{0.35in}\fullmoon~~Prefer not to answer\\

During a typical week how many different places do you go where you connect to a wifi network (e.g. home, work, a public library, etc.)?\\
\\\hspace*{0.35in}\fullmoon~~0-1
\\\hspace*{0.35in}\fullmoon~~2-3
\\\hspace*{0.35in}\fullmoon~~4-5
\\\hspace*{0.35in}\fullmoon~~5+
\\\hspace*{0.35in}\fullmoon~~Don't know
\\\hspace*{0.35in}\fullmoon~~Prefer not to answer\\

\subsubsection{Participant Knowledge of DNS}
Before this survey have you ever heard of DNS?\\
You do not need to have any prior knowledge of DNS to take part in this survey.\\
\\\hspace*{0.35in}\fullmoon~~Yes, I have heard of DNS and I know what it does
\\\hspace*{0.35in}\fullmoon~~Yes, I have heard of DNS, but I do not know what it does
\\\hspace*{0.35in}\fullmoon~~No, I have not heard of DNS\\
\emph{If yes,} What do you think DNS does?\\
\subsubsection{Participants Interact with Browser Interface}
\begin{figure}[t!]
  \includegraphics[width=0.6\linewidth]{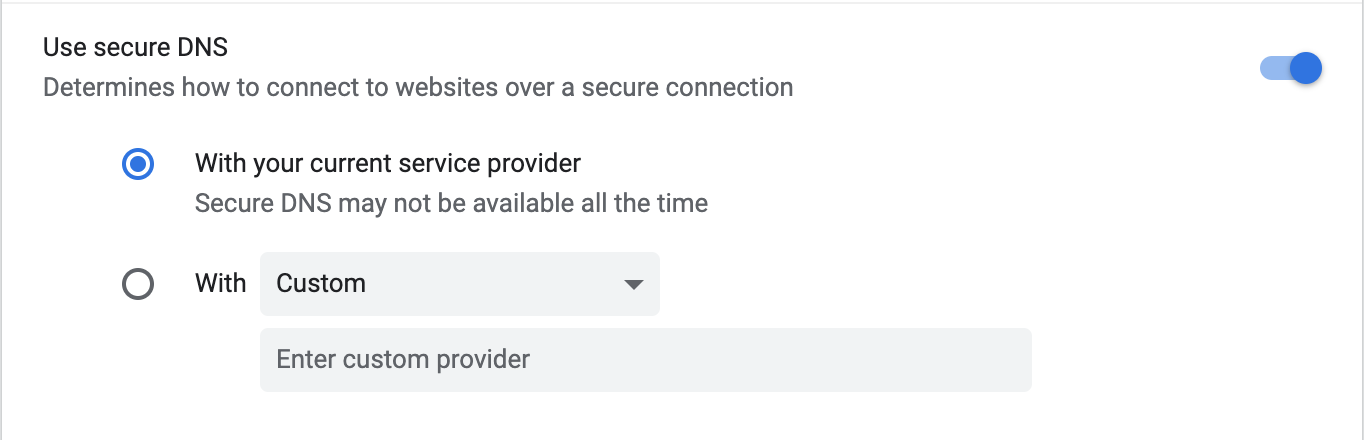}
  \caption{Chrome Browser Interface}
  \label{fig:Chrome_Interface}
\end{figure}

On the next page, you will shown a network setting that you could see in a browser on your laptop or desktop computer.\\
Please select the setting option that you would choose if you encountered this setting in your own browser.\\
The next button will appear after a few seconds to give you time to choose a setting option.\\
Please note that this page will not modify the settings in your browser.\\
\emph{Figure~\ref{fig:Chrome_Interface} shows the interface that a participant would be shown.}\\
\emph{Here, we only show the questions asked for the Chrome interface, but participants were shown either a Chrome, Edge, Firefox, Opera, or Brave interface.} 

 \emph{If secure DNS disabled,} Why did you choose to disable secure DNS?\\

 \emph{If secure DNS enabled,} Why did you choose to enable secure DNS?\\

 \emph{If current service provider chosen as resolver,} Why did you choose "your current service provider" as your resolver?\\

 \emph{If resolver from dropdown chosen,} Why did you choose \{resolver name\} as your resolver?\\

 \emph{If custom resolver chosen,} Why did you choose to use a custom resolver?\\

 \emph{If custom resolver chosen,} Why did you choose \{custom resolver name\} as your custom resolver?\\
\subsubsection{General Questions about Setting Choices}
How certain are you that the setting you chose will allow you to access the Internet as you usually do?\\
\\\hspace*{0.35in}\fullmoon~~Very certain
\\\hspace*{0.35in}\fullmoon~~Somewhat certain
\\\hspace*{0.35in}\fullmoon~~Neither certain nor uncertain
\\\hspace*{0.35in}\fullmoon~~Somewhat uncertain
\\\hspace*{0.35in}\fullmoon~~Very uncertain\\

Why did you not choose any of the other setting options?\\

Regardless of your knowledge about this setting, what advantages do you think it might have?\\

Regardless of your knowledge about this setting, what disadvantages do you think it might have?\\

While you were choosing a setting, what additional information, if any, would you have wanted?\\

If you wanted to make a more informed choice about this setting where would you go to learn more? Please select all that apply\\
\\\hspace*{0.35in}\fullmoon~~Social media, e.g., Twitter or Facebook
\\\hspace*{0.35in}\fullmoon~~A security blog
\\\hspace*{0.35in}\fullmoon~~Ask a friend, family member, or coworker Look up the setting online
\\\hspace*{0.35in}\fullmoon~~A news outlet (e.g., TV, online)
\\\hspace*{0.35in}\fullmoon~~Reddit or other online forums
\\\hspace*{0.35in}\fullmoon~~Company website for your browser
\\\hspace*{0.35in}\fullmoon~~Other\\
\begin{figure}[!ht]
  \includegraphics[width=0.6\linewidth]{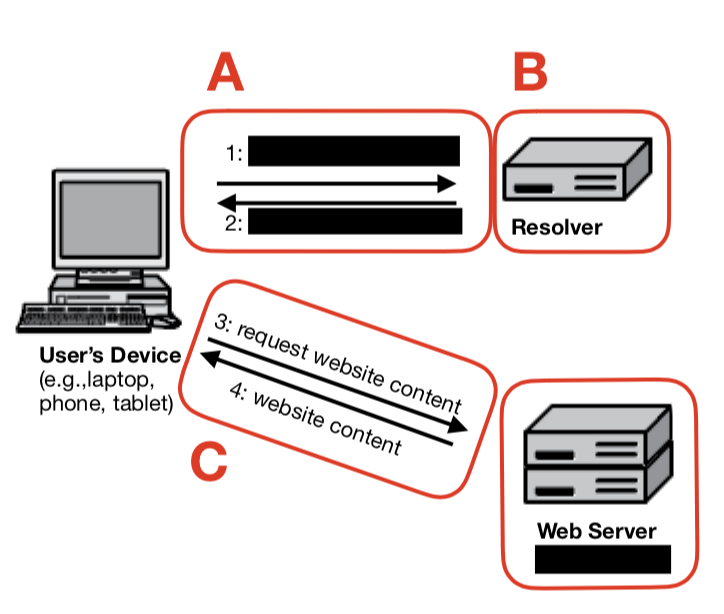}
  \caption{Diagram of DNS}
  \label{fig:DNS_Diagram_Regions}
\end{figure}

Referring to the diagram above (Figure~\ref{fig:DNS_Diagram_Regions}), at which places do you think your Internet Service Provider (the company that you pay for Internet access in your home) could know what websites you are visiting if the setting you saw on the last page was enabled and queries were sent to a single DNS resolver (regardless of what you actually chose)?\\
Please select all that apply.\\
If you are uncertain please just make your best guess.\\
\\\hspace*{0.35in}\fullmoon~~A
\\\hspace*{0.35in}\fullmoon~~B
\\\hspace*{0.35in}\fullmoon~~C
\\\hspace*{0.35in}\fullmoon~~D
\\\hspace*{0.35in}\fullmoon~~None of the above
\\\hspace*{0.35in}\fullmoon~~Don't know\\

Referring to the diagram above, at which places do you think a third party (someone not directly involved in you reaching a website) could know what websites you are visiting if the setting you saw on the last page\\
was enabled and queries were sent to a single DNS resolver (regardless of what you actually chose)?\\
Please select all that apply.\\
If you are uncertain please just make your best guess.\\
\\\hspace*{0.35in}\fullmoon~~A
\\\hspace*{0.35in}\fullmoon~~B
\\\hspace*{0.35in}\fullmoon~~C
\\\hspace*{0.35in}\fullmoon~~D
\\\hspace*{0.35in}\fullmoon~~None of the above 
\\\hspace*{0.35in}\fullmoon~~Don't know\\
\subsubsection{Participants Interact with Mobile Interface}
\begin{figure}[!ht]
  \includegraphics[width=0.6\linewidth]{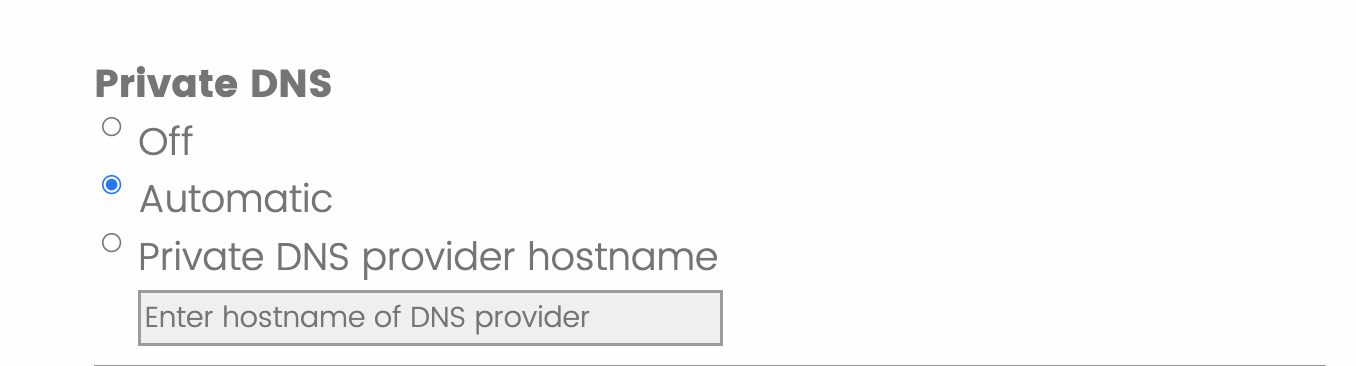}
  \caption{Mobile Interface}
  \label{fig:mobile_custom}
\end{figure}

On the next page, you will be shown a network setting that you could see on a mobile phone.\\
Please select the setting option that you would choose if you encountered this setting in your own browser.\\
The next button will appear after a few seconds to give you time to choose a setting option.\\
Please note that this page will not modify the settings on your phone.\\
\emph{Figure~\ref{fig:mobile_custom} shows the interface that a participant would be shown.}\\

 \emph{If resolver from dropdown chosen,} Why did you select the \{resolver name\} setting for Private DNS?\\

 \emph{If custom resolver chosen,} Why did you enter \{custom resolver name\} as your DNS Provider hostname?\\
\subsubsection{DNS Explained}
\begin{figure}[t!]
  \includegraphics[width=0.6\linewidth]{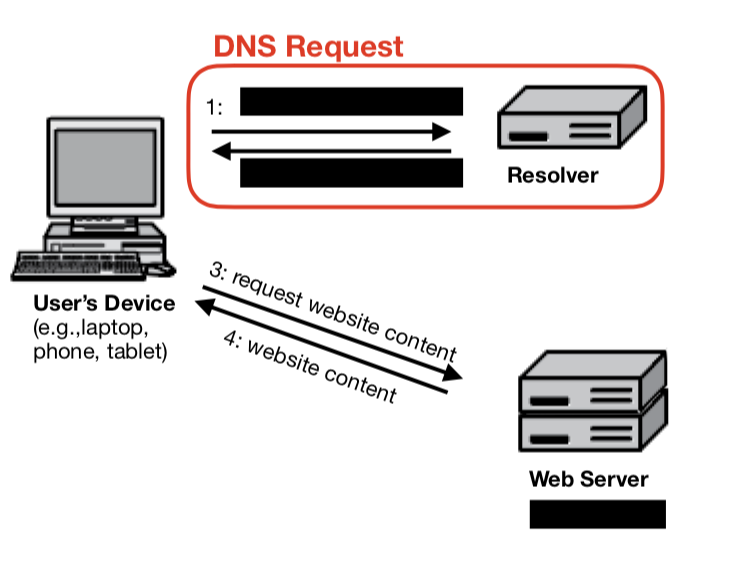}
  \caption{Diagram of DNS}
  \label{fig:DNS_Diagram_No_Regions}
\end{figure}
DNS stands for Domain Name System.\\
It is the system that takes a human-readable name like \censor{noise.cs.uchicago.edu} and returns an Internet Protocol (IP) address, like \censor{128.135.24.19}.\\

Whenever your browser tries to contact a website (when you click on a link, when your browser loads an image, etc.), your browser sends a request over the Internet to a resolver to look up the IP address for that website allowing your computer find that site or resource.\\
A DNS request is like looking up the street address (e.g., \censor{5801 S Ellis Ave. Chicago, IL 60637}) of a particular location (e.g., \censor{The University of Chicago}).\\
Either can be used to refer to the place, but only the street address can be used to send a package.\\
Similarly, data on the Internet is sent between IP addresses (not between names).\\
Typically, DNS requests are not encrypted.\\
This means that your Internet Service Provider or any third-party on the path between your computer, laptop, phone, etc. and the other side of the communication may be able to see which websites you are about to access and potentially change what is returned.\\

What does DNS do?\\
\\\hspace*{0.35in}\fullmoon~~Encrypts all network traffic
\\\hspace*{0.35in}\fullmoon~~Optimizes the order of your search results
\\\hspace*{0.35in}\fullmoon~~Maps a website to an IP address\\
Normally these DNS requests are made in plain text, making it easy for your Internet Service Provider or third- parties to see what website you’re about to access.\\
When encrypted DNS is used, the request is encrypted, meaning that third parties and your Internet Service Provider can't see what website you are trying to access at this point.\\
However, encrypted DNS can be incompatible with parental controls and malware detectors, can make browsing the internet slower, and the place that translates the URL to an IP address (resolver) knows what site you are trying to reach.\\
\begin{figure}[t!]
  \includegraphics[width=0.6\linewidth]{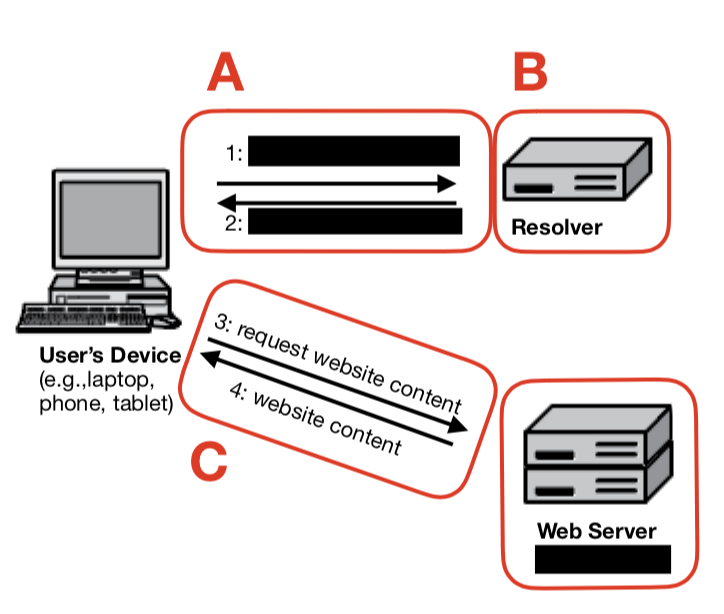}
  \caption{Diagram of DNS}
  \label{fig:DNS_Diagram_Regions1}
\end{figure}
Referring to the diagram above (Figure~\ref{fig:DNS_Diagram_No_Regions}), at which places do you think your Internet Service Provider (the company that you pay for Internet access in your home) could know what websites you are visiting if the setting you saw on the last page was enabled and queries were sent to a single DNS resolver (regardless of what you actually chose)?\\
Please select all that apply.\\
If you are uncertain please just make your best guess.\\
\\\hspace*{0.35in}\fullmoon~~A
\\\hspace*{0.35in}\fullmoon~~B
\\\hspace*{0.35in}\fullmoon~~C
\\\hspace*{0.35in}\fullmoon~~D
\\\hspace*{0.35in}\fullmoon~~None of the above
\\\hspace*{0.35in}\fullmoon~~Don't know\\

Referring to the diagram above (Figure~\ref{fig:DNS_Diagram_No_Regions}), at which places do you think a third party (someone not directly involved in you reaching a website) could know what websites you are visiting if the setting you saw on the last page was enabled and queries were sent to a single DNS resolver (regardless of what you actually chose)?\\
Please select all that apply.\\
If you are uncertain please just make your best guess.\\
\\\hspace*{0.35in}\fullmoon~~A
\\\hspace*{0.35in}\fullmoon~~B
\\\hspace*{0.35in}\fullmoon~~C
\\\hspace*{0.35in}\fullmoon~~D
\\\hspace*{0.35in}\fullmoon~~None of the above
\\\hspace*{0.35in}\fullmoon~~Don't know\\

Now that you have read about encrypted DNS, when it is enabled, how safe do you think while you are browsing the Internet?\\
\\\hspace*{0.35in}\fullmoon~~Very safe
\\\hspace*{0.35in}\fullmoon~~Somewhat safe
\\\hspace*{0.35in}\fullmoon~~Neither safe or unsafe
\\\hspace*{0.35in}\fullmoon~~Somewhat unsafe
\\\hspace*{0.35in}\fullmoon~~Very unsafe\\

Why?\\

\subsubsection{Participants Interact with Browser Interface Again}
\begin{figure}[t!]
  \includegraphics[width=0.6\linewidth]{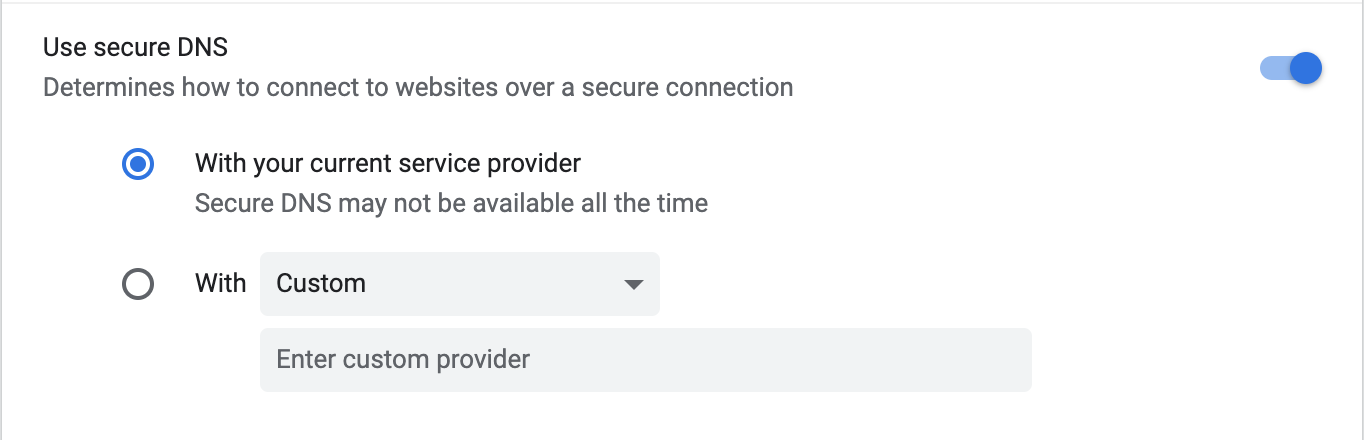}
  \caption{Chrome Browser Interface}
  \label{fig:Chrome_Interface2}
\end{figure}

On the next page, you will be shown the same network setting that you saw earlier in the survey.\\
Please select the setting option that you would choose if you encountered this setting in your own browser on your laptop or desktop computer given what you have read here today.\\
The next button will appear after a few seconds to give you time to choose a setting option.\\
Please note that this page will not modify the settings in your browser.\\
\emph{Figure~\ref{fig:Chrome_Interface2} shows the interface that a participant would be shown.}\\

 \emph{If secure DNS disabled,} Why did you choose to disable secure DNS?\\

 \emph{If secure DNS enabled,} Why did you choose to enable secure DNS?\\

 \emph{If current service provider chosen as resolver,} Why did you choose "your current service provider" as your resolver?\\

 \emph{If resolver from dropdown is chosen,} Why did you choose \{resolver name\} as your resolver?\\

 \emph{If custom resolver is chosen,} Why did you choose to use a custom resolver?\\

Why did you choose \{custom resolver name\} as your custom resolver?\\

\subsubsection{Questions about Setting Choices}

Why did you not choose any of the other setting options?\\

How certain are you that the setting you chose will allow you to access the Internet as you usually do?\\
\\\hspace*{0.35in}\fullmoon~~Very certain
\\\hspace*{0.35in}\fullmoon~~Somewhat certain
\\\hspace*{0.35in}\fullmoon~~Neither certain nor uncertain 
\\\hspace*{0.35in}\fullmoon~~Somewhat uncertain
\\\hspace*{0.35in}\fullmoon~~Very uncertain\\

\subsubsection{Encrypted DNS Preferences}

Prior to this survey, had you heard of encrypted DNS? You may have also heard it referred to as referred to as DNS over HTTPS (DoH), DNS over TLS (DoT), or Secure DNS.\\
\\\hspace*{0.35in}\fullmoon~~Yes
\\\hspace*{0.35in}\fullmoon~~No/Don't know\\

Where did you hear about encrypted DNS?\\
\\\hspace*{0.35in}\fullmoon~~A friend, family member, or coworker 
\\\hspace*{0.35in}\fullmoon~~A security blog
\\\hspace*{0.35in}\fullmoon~~Social media, e.g., Twitter or Facebook 
\\\hspace*{0.35in}\fullmoon~~Company website for your browser 
\\\hspace*{0.35in}\fullmoon~~Reddit or other online forums
\\\hspace*{0.35in}\fullmoon~~A news outlet (e.g. TV, online)
\\\hspace*{0.35in}\fullmoon~~Other\\
\begin{figure}[t!]
  \includegraphics[width=0.6\linewidth]{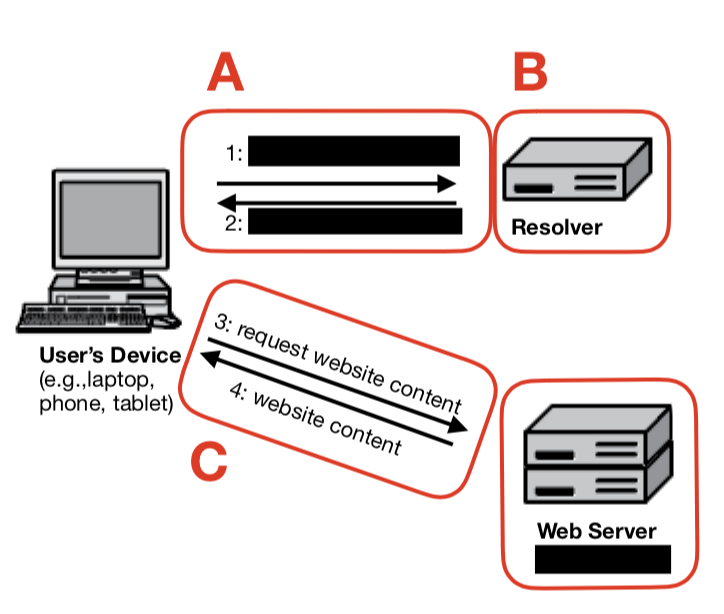}
  \caption{Diagram of DNS}
  \label{fig:DNS_Diagram_Regions2}
\end{figure}
Sometimes companies that host websites also run resolvers (e.g., B and D in the diagram above (Figure~\ref{fig:DNS_Diagram_Regions2}) are run by the same company).\\
Please select the option that most closely reflects your preference.\\
\\\hspace*{0.35in}\fullmoon~~I would prefer that B and D were run by the same company.
\\\hspace*{0.35in}\fullmoon~~I would prefer that B and D were run by the different companies.
\\\hspace*{0.35in}\fullmoon~~Other
\\\hspace*{0.35in}\fullmoon~~No preference\\

There are many different companies that operate resolvers.\\
Please select the option that most closely reflects your preference when the companies that host websites you are trying to reach does not run a resolver.\\
\\\hspace*{0.35in}\fullmoon~~I would prefer that my computer always contact the same resolver (B) who would see all of my DNS requests.
\\\hspace*{0.35in}\fullmoon~~I would prefer that my computer distributes my requests over multiple resolvers (B) who would each see some of my DNS requests.
\\\hspace*{0.35in}\fullmoon~~Other
\\\hspace*{0.35in}\fullmoon~~Don't know\\

In the case that your chosen resolver does not support encrypted DNS or your encrypted DNS requests fail, please select the behavior that you would prefer.\\
\\\hspace*{0.35in}\fullmoon~~I would prefer to have my computer make an unencrypted query so the page can load.
\\\hspace*{0.35in}\fullmoon~~I would prefer to have my computer fail to reach the site/load the resource.
\\\hspace*{0.35in}\fullmoon~~Other
\\\hspace*{0.35in}\fullmoon~~No preference\\

Do you think this a setting you should be able to modify on your phone or computer browser?\\
\\\hspace*{0.35in}\fullmoon~~Yes
\\\hspace*{0.35in}\fullmoon~~No
\\\hspace*{0.35in}\fullmoon~~Don't know\\

Why?\\

\subsubsection{Trust in Specific DNS Providers}

\emph{Here, we only show the questions asked for Quad9, but participants were asked about either Quad9, Google, Cloudflare, NextDNS, or Comcast.}
Had you heard of Quad9 prior to this survey?\\
\\\hspace*{0.35in}\fullmoon~~Yes 
\\\hspace*{0.35in}\fullmoon~~Maybe 
\\\hspace*{0.35in}\fullmoon~~No\\

Please select the answer option that most accurately describes your agreement with the following statement.\\ 
 I trust Quad9?\\
\\\hspace*{0.35in}\fullmoon~~Strongly agree
\\\hspace*{0.35in}\fullmoon~~Somewhat agree
\\\hspace*{0.35in}\fullmoon~~Neither agree nor disagree 
\\\hspace*{0.35in}\fullmoon~~Somewhat disagree 
\\\hspace*{0.35in}\fullmoon~~ Strongly disagree\\

\subsubsection{Trust in Providers and Browsers}
Please select the answer option that most accurately describes your agreement with the following statement. \\
I trust my Internet Service Provider (e.g., Xfinity, Time Warner, Spectrum)?\\
\\\hspace*{0.35in}\fullmoon~~Strongly agree
\\\hspace*{0.35in}\fullmoon~~Somewhat agree
\\\hspace*{0.35in}\fullmoon~~Neither agree nor disagree
\\\hspace*{0.35in}\fullmoon~~Somewhat disagree
\\\hspace*{0.35in}\fullmoon~~Strongly disagree
\\\hspace*{0.35in}\fullmoon~~I don't know who my Internet Service Provider is\\

Please select the answer option that most accurately describes your agreement with the following statement. \\
I trust my mobile service provider (e.g., AT\&T, Verizon, etc.)?\\
\\\hspace*{0.35in}\fullmoon~~Strongly agree
\\\hspace*{0.35in}\fullmoon~~Somewhat agree
\\\hspace*{0.35in}\fullmoon~~Neither agree nor disagree
\\\hspace*{0.35in}\fullmoon~~Somewhat disagree
\\\hspace*{0.35in}\fullmoon~~Strongly disagree
\\\hspace*{0.35in}\fullmoon~~I don't know who my mobile service provider is\\

Please select the answer option that most accurately describes your agreement with the following statement.\\
I trust my primary web browser (e.g., Chrome, Firefox, Safari)?\\
\\\hspace*{0.35in}\fullmoon~~Strongly agree
\\\hspace*{0.35in}\fullmoon~~Somewhat agree
\\\hspace*{0.35in}\fullmoon~~Neither agree nor disagree 
\\\hspace*{0.35in}\fullmoon~~Somewhat disagree 
\\\hspace*{0.35in}\fullmoon~~Strongly disagree\\

\subsubsection{Access Interfaces through Actual Browsers}
Which of the following browsers are you currently able to access on the device you are using to take this survey?\\
\\\hspace*{0.35in}\fullmoon~~Opera
\\\hspace*{0.35in}\fullmoon~~Chrome
\\\hspace*{0.35in}\fullmoon~~Microsoft Edge Brave
\\\hspace*{0.35in}\fullmoon~~Firefox
\\\hspace*{0.35in}\fullmoon~~None of the above\\
\emph{Here, we only show the questions asked for the Chrome interface, but participants were asked about either Chrome, Edge, Firefox, Opera, or Brave.} 
Please open a new Chrome browser window and navigate to your browser's settings page by following the written instructions or video below:\\
At the top right of the window, click More\\
Click Settings\\
Under Privacy and security select Security\\
\emph{Please note that these instructions may not be compatible with old versions of Chrome.}

 \emph{\{Video with instructions shown below\} }\\

Were you able to reach this page?
\\\hspace*{0.35in}\fullmoon~~Yes
\\\hspace*{0.35in}\fullmoon~~No\\

Can you find "Use secure DNS" on this page?
\emph{If you are using an old version of Chrome the secure DNS setting might not be available.}
\\\hspace*{0.35in}\fullmoon~~Yes
\\\hspace*{0.35in}\fullmoon~~No
Near the bottom of this page you should see an option called ''Use secure DNS''. Is this setting enabled?
\\\hspace*{0.35in}\fullmoon~~Yes
\\\hspace*{0.35in}\fullmoon~~No
\\\hspace*{0.35in}\fullmoon~~Don't know\\

 \emph{If yes,} Which setting is selected under ''Use secure DNS?''
\\\hspace*{0.35in}\fullmoon~~With your current service provider
\\\hspace*{0.35in}\fullmoon~~With ---\\

 \emph{If "With ---" is selected,} What DNS provider is selected?
\\\hspace*{0.35in}\fullmoon~~Custom
\\\hspace*{0.35in}\fullmoon~~CleanBrowsing (Family Filter)
\\\hspace*{0.35in}\fullmoon~~OpenDNS
\\\hspace*{0.35in}\fullmoon~~Quad9 (9.9.9.9)
\\\hspace*{0.35in}\fullmoon~~Cloudflare (1.1.1.1)
\\\hspace*{0.35in}\fullmoon~~NextDNS
\\\hspace*{0.35in}\fullmoon~~Google (Public DNS)\\

Do you remember seeing the "secure DNS" setting in Chrome prior to this survey?
\\\hspace*{0.35in}\fullmoon~~Yes
\\\hspace*{0.35in}\fullmoon~~No
\\\hspace*{0.35in}\fullmoon~~Don't Know\\

 \emph{If yes,} Do you remember changing the "secure DNS" setting in Chrome?
\\\hspace*{0.35in}\fullmoon~~Yes
\\\hspace*{0.35in}\fullmoon~~No
\\\hspace*{0.35in}\fullmoon~~Don't Know\\

 \emph{If yes,} Why did you decide to change the "secure DNS" setting in Chrome?\\

 \emph{If no,} Why did you decide to not change, the "secure DNS" setting in Chrome?
\subsubsection{Access Interfaces through Mobile Device}
What is the operating system of your primary smartphone?\\
\\\hspace*{0.35in}\fullmoon~~Android
\\\hspace*{0.35in}\fullmoon~~iOS (Apple device)
\\\hspace*{0.35in}\fullmoon~~Other
\\\hspace*{0.35in}\fullmoon~~I do not use a smartphone\\

What version of iOS does your primary smartphone run?\\
The version of iOS can be found by going to Settings > General > About on your phone.\\

What version of Android does your primary smartphone run?\\
The version of Android can be found by going to Settings > System > Advanced > System update on your phone.\\
\emph{Here, we only show the questions asked for Android devices, but participants were asked about either iOS or Android.}
Open your phone's Settings app.\\
Tap Network \& internet > Advanced > Private DNS.\\

 Were you able to make it to this page?\\
\\\hspace*{0.35in}\fullmoon~~Yes
\\\hspace*{0.35in}\fullmoon~~No\\
What is you current setting on this page?\\
\\\hspace*{0.35in}\fullmoon~~Off
\\\hspace*{0.35in}\fullmoon~~Automatic
\\\hspace*{0.35in}\fullmoon~~Private DNS Provider hostname\\

 Do you remember visiting this page before now?\\
\\\hspace*{0.35in}\fullmoon~~Yes
\\\hspace*{0.35in}\fullmoon~~No/Don't know\\

Do you remember changing this setting?\\
\\\hspace*{0.35in}\fullmoon~~Yes
\\\hspace*{0.35in}\fullmoon~~No/Don't know\\

If you changed the setting or not, why did you make that decision?\\
\subsubsection{Demographics}
What is your age?\\
\\\hspace*{0.35in}\fullmoon~~18 - 24 
\\\hspace*{0.35in}\fullmoon~~25 - 34
\\\hspace*{0.35in}\fullmoon~~35 - 44
\\\hspace*{0.35in}\fullmoon~~45 - 54
\\\hspace*{0.35in}\fullmoon~~55 - 64
\\\hspace*{0.35in}\fullmoon~~65 - 74
\\\hspace*{0.35in}\fullmoon~~75 - 84
\\\hspace*{0.35in}\fullmoon~~85 or older
\\\hspace*{0.35in}\fullmoon~~Prefer not to answer\\

What is your gender?\\
\\\hspace*{0.35in}\fullmoon~~Woman 
\\\hspace*{0.35in}\fullmoon~~Man 
\\\hspace*{0.35in}\fullmoon~~Non-Binary
\\\hspace*{0.35in}\fullmoon~~Prefer not to disclose
\\\hspace*{0.35in}\fullmoon~~Prefer to self-describe\\

What is the highest level of school you have completed or the highest degree you have received?\\
\\\hspace*{0.35in}\fullmoon~~Less than high school 
\\\hspace*{0.35in}\fullmoon~~High school graduate 
\\\hspace*{0.35in}\fullmoon~~Some college
\\\hspace*{0.35in}\fullmoon~~2 year degree
\\\hspace*{0.35in}\fullmoon~~4 year degree 
\\\hspace*{0.35in}\fullmoon~~Professional degree 
\\\hspace*{0.35in}\fullmoon~~Doctorate
\\\hspace*{0.35in}\fullmoon~~Prefer not to answer\\

Which of the following best describes your educational background or job field?\\
\\\hspace*{0.35in}\fullmoon~~I have an education in, or work in, the field of computer science, computer engineering or IT
\\\hspace*{0.35in}\fullmoon~~I DO NOT have an education in, nor do I work in, the field of computer science, computer engineering or IT
\\\hspace*{0.35in}\fullmoon~~Prefer not to answer\\

\subsubsection{Feedback}
If you have any further feedback, questions, comments, concerns, or anything else you want to tell us, please leave a comment below.

\end{document}